\documentclass[twocolumn,aps,superscriptaddress]{revtex4-2}
\usepackage{hyperref}
\usepackage[table,dvipsnames]{xcolor}
\usepackage{physics}
\usepackage{amsmath}
\usepackage{amssymb}
\usepackage{comment}
\usepackage{algpseudocode}
\usepackage{subfigure}
\usepackage{graphicx}
\usepackage{qcircuit}
\usepackage{float}
\usepackage[section]{placeins}
\colorlet{gray1}{gray!40}
\colorlet{gray2}{gray!30}
\colorlet{gray3}{gray!20}
\colorlet{gray4}{gray!10}
\newcommand{\qwxdotted}[1][-1]{\ar @{..} [#1,0] \ar @{..} [#1,0] \ar @{..} [#1,0] \ar @{..} [#1,0] \ar @{..} [#1,0] \ar @{..} [#1,0] \ar @{..} [#1,0] \ar @{..} [#1,0] \ar @{..} [#1,0] \ar @{..} [#1,0] \ar @{..} [#1,0] \ar @{..} [#1,0]}

\begin{document}

\title{Co-Designed Adaptive Quantum State Preparation Protocols}

\author{Mafalda Ramôa}
\email{mafalda@vt.edu}
\affiliation{Department of Physics, Virginia Tech, Blacksburg, VA, 24061, USA}
\affiliation{Virginia Tech Center for Quantum Information Science and Engineering, Blacksburg, VA 24061, USA}
\affiliation{International Iberian Nanotechnology Laboratory (INL), Portugal}
\affiliation{High-Assurance Software Laboratory (HASLab), Portugal}
\affiliation{Department of Computer Science, University of Minho, Portugal}

\author{Luis Paulo Santos}
\affiliation{International Iberian Nanotechnology Laboratory (INL), Portugal}
\affiliation{High-Assurance Software Laboratory (HASLab), Portugal}
\affiliation{Department of Computer Science, University of Minho, Portugal}

\author{Nicholas J. Mayhall}
\affiliation{Virginia Tech Center for Quantum Information Science and Engineering, Blacksburg, VA 24061, USA}
\affiliation{Department of Chemistry, Virginia Tech, Blacksburg, VA, 24061, USA}

\author{Edwin Barnes}
\affiliation{Department of Physics, Virginia Tech, Blacksburg, VA, 24061, USA}
\affiliation{Virginia Tech Center for Quantum Information Science and Engineering, Blacksburg, VA 24061, USA}

\author{Sophia E. Economou}
\affiliation{Department of Physics, Virginia Tech, Blacksburg, VA, 24061, USA}
\affiliation{Virginia Tech Center for Quantum Information Science and Engineering, Blacksburg, VA 24061, USA}

\begin{abstract}
We propose a co-designed variant of ADAPT-VQE (Co-ADAPT-VQE) where the quantum hardware is taken into account in the construction of the ansatz. This framework can be readily used to optimize state preparation circuits for any device, addressing shortcomings such as limited connectivity, short coherence times, and variable gate errors. We exemplify the impact of Co-ADAPT-VQE by creating state preparation circuits for devices with linear nearest-neighbor (LNN) connectivity. We show a reduction of the \texttt{CNOT} count of the final circuits by up to 97\% for 12-14 qubit systems, with the impact being greater for larger and more strongly correlated systems. Surprisingly, the circuits created by Co-ADAPT-VQE provide an over 70\% \texttt{CNOT} count reduction with respect to the original ADAPT-VQE in all-to-all connectivity, despite being restricted to LNN qubit interactions.

\end{abstract}

\maketitle

\section{Introduction}

In the near term, the potential of quantum computing is limited by hardware, as devices are unable to execute long coherent sequences of gates \cite{preskillQuantumComputingNISQ2018}. This has prevented experimental demonstrations of the earliest proposals of quantum algorithms, such as Shor's factoring algorithm \cite{shorPolynomialTimeAlgorithmsPrime1996} or Grover's search algorithm \cite{groverFastQuantumMechanical1996}, except for very small instances. While these algorithms have the advantage of offering provable speedups, they require very deep circuits, which remain out of reach before the fault-tolerant quantum computing (FTQC) era.

In light of this misalliance between available hardware and algorithms, the Variational Quantum Eigensolver (VQE) was proposed in 2014 as a near-term option for ground state preparation \cite{peruzzoVariationalEigenvalueSolver2014}. The field of variational quantum algorithms (VQAs) has since grown dramatically, with proposals covering virtually all realms of application, including quantum chemistry, combinatorial optimization, condensed matter, factoring, and machine learning \cite{cerezoVariationalQuantumAlgorithms2021a,farhiQuantumApproximateOptimization2014,cadeStrategiesSolvingFermiHubbard2020,anschuetzVariationalQuantumFactoring2018,schuldCircuitcentricQuantumClassifiers2020}. VQAs create a feedback loop between a classical optimizer and a quantum computer, with the role of the latter being restricted to state preparation (via a parameterized circuit, or \textit{ansatz}) and measurement. This allows the computation to occur iteratively---with the quantum computer measuring the cost function and the classical computer updating the quantum circuit parameters---, which reduces the depth of the circuits that must be executed coherently.

Several types of ans\"atze have been proposed in the literature, which can be broadly classified as \textit{hardware-tailored} or \textit{problem-tailored}. The former are chosen to include only gates and interactions which are natively available in the target device, while the latter are based on knowledge about the problem at hand; they may, e.g., preserve system symmetries or draw inspiration from previous classical methods. An example in the former category is the hardware-efficient ansatz (HEA) \cite{kandalaHardwareefficientVariationalQuantum2017}, which offers the alluring feature of producing circuits that are ready to be executed on hardware, such that no transpilation overheads are incurred. However, this ansatz has been shown to suffer from trainability issues, known as \textit{barren plateaus}, that prevent it from scaling efficiently with the system size \cite{mccleanBarrenPlateausQuantum2018}. In the second category we have the adaptive derivative-assembled problem-tailored (ADAPT)-VQE algorithm \cite{grimsleyAdaptiveVariationalAlgorithm2019}, which is not only problem- but also system-tailored. In this case, the structure of the ansatz is dictated by on-the-fly measurements on the quantum computer that aim to maximize the contribution of each ansatz element towards lowering the cost function. The shallow circuits offered by ADAPT-VQE, combined with its remarkable resilience against barren plateaus and local minima \cite{grimsleyADAPTVQEInsensitiveRough2022a}, has attracted the interest of the research community, leading to a substantial body of follow-up studies. Several improvements have been proposed for further optimizing the circuits \cite{ramoaReducingResourcesRequired2024,yordanovQubitexcitationbasedAdaptiveVariational2021,tangQubitADAPTVQEAdaptiveAlgorithm2021,anastasiouTETRISADAPTVQEAdaptiveAlgorithm2022}, reducing measurement costs \cite{ramoaReducingMeasurementCosts2024b,anastasiouHowReallyMeasure2023}, and generalizing the algorithm to applications such as thermal states, lattice models, periodically driven systems, or combinatorial optimization \cite{sambasivamTEPIDADAPTAdaptiveVariational2025,dykeScalingAdaptiveQuantum2023,kumarFloquetADAPTVQEQuantumAlgorithm2025,zhuAdaptiveQuantumApproximate2022}. However, in contrast with hardware-efficient ans\"atze, ADAPT-VQE, in its original formulation, is completely agnostic to the target hardware, which results in high transpilation overheads.

\begin{table*}
\begin{center}
\begin{tabular}{ c | c | c | c |}
\cline{2-4}
& \multicolumn{3}{|c|}{CNOT Count for Chemical Accuracy} \\
\cline{2-4}
& ATA Connectivity & LNN Connectivity - Previous & LNN Connectivity - This Work \\ 
\hline
\multicolumn{1}{|c|}{Linear H$_6$} & 854 & 2945 & 348 (12\%) \\
\hline  
\multicolumn{1}{|c|}{Linear H$_7$} & 437 & 1497 & 92 (6\%) \\
\hline 
\multicolumn{1}{|c|}{Triangular H$_6$} & 1533 & 5323 & 184 (3\%) \\
\hline
\end{tabular}
\end{center}
\caption{Comparison of the \texttt{CNOT} counts required by CEO-ADAPT-VQE \cite{ramoaReducingResourcesRequired2024} to reach chemical accuracy in all-to-all (ATA) and linear nearest-neighbor (LNN) connectivity. For the latter, we consider the previous state-of-the-art (obtained from the ATA circuit using the Qiskit (version 2.2.1) transpiler set to the maximum optimization level \cite{QiskitCommunity2017}) as well as an instance of Co-ADAPT-VQE specifically tailored to reduce the \texttt{CNOT} count of the final circuits.  The percentages in parentheses represent the ratio of the \texttt{CNOT} count of our method to the \texttt{CNOT} count of the previous one. Three different systems are considered: linear H$_6$ (12 qubits), linear H$_7$ (14 qubits), and triangular H$_6$ (12 qubits).}
\label{tab:h_counts}
\end{table*}

In this work, we propose to combine the benefits of hardware- and problem-tailored ans\"atze by co-designing ADAPT-VQE such that it takes into account architecture restrictions and device limitations. Our proposed algorithm, Co-ADAPT-VQE, modifies the selection criterion to penalize circuit components which are less suitable for the underlying quantum computer. This customizable penalty may be based on generic features, such as the depth or two-qubit gate count of the transpiled circuit, or on device-specific features, such as gate errors. This allows the creation of more compact and noise-resilient circuits using a protocol that is simultaneously hardware-, problem-, and system-tailored. Our algorithm maintains the optimization advantages of ADAPT-VQE while augmenting it with all the benefits stemming from hardware efficiency.

We exemplify the potential of Co-ADAPT-VQE with a standard task in quantum simulation: finding molecular ground states using state preparation circuits. We consider a restriction to linear nearest-neighbor (LNN) connectivity and aim to minimize the number of two-qubit gates. Table \ref{tab:h_counts} shows that the two-qubit gate counts suffer an over 3-fold increase under transpilation from ATA to LNN connectivity using state-of-the-art methods; yet, Co-ADAPT-VQE is able to reduce the LNN counts by up to 97\%, resulting in an \textit{improvement} over the ATA connectivity circuit, despite the additional interaction restrictions. 

The rest of this paper is organized as follows. In Sec.~\ref{s:background}, we introduce relevant background, covering the algorithms, operators and circuits that are used in this work. Section \ref{s:co-adapt} introduces Co-ADAPT-VQE and features an example proposal where it is used to minimize the LNN \texttt{CNOT} count of the leading variants of ADAPT-VQE. The results of this proposal are presented and discussed in Sec.~\ref{s:results}. Concluding remarks are contained in Sec.~\ref{s:conclusion}.

\section{Background}
\label{s:background}

In this section, we provide an overview of the topics necessary for understanding the contents of the paper. We introduce the VQE and ADAPT-VQE algorithms, as well as the circuit implementations of the corresponding ansatz operators. Readers familiar with this background may skip this section.

\subsection{VQE}
\label{ss:vqe}

The VQE \cite{peruzzoVariationalEigenvalueSolver2014} was proposed as a hybrid quantum-classical algorithm for the electronic structure problem. Its purpose is to find the eigenstates of the electronic structure Hamiltonian $\mathcal{H}_e$, i.e., solutions of the time-independent Schr\"odinger equation $\mathcal{H}_e\ket{\psi}=E\ket{\psi}$. It is common to focus on ground state preparation, although the method has been generalized to excited states as well \cite{asthanaQuantumSelfconsistentEquationofmotion2023,yordanovQubitexcitationbasedAdaptiveVariational2021,higgottVariationalQuantumComputation2019}.

\begin{figure}[htbp]
    \centering
    \includegraphics[width=\columnwidth]{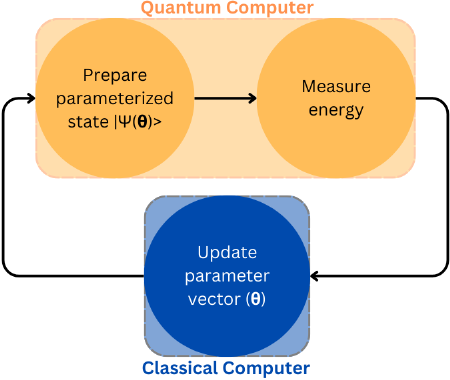}
    \caption{Flowchart for the VQE algorithm.}
    \label{fig:vqe}
\end{figure}

The steps of the algorithm are summarized in Fig.~\ref{fig:vqe}. The quantum computer prepares parameterized states $\ket{\psi(\vec{\theta})}$ and measures the corresponding energy, while the classical computer updates the parameter vector $\vec{\theta}$ to minimize this energy. The parameters are updated iteratively until a convergence criterion is met (e.g., the energy change from a parameter update falls below a given threshold). 

The exact form of the ansatz $\ket{\psi(\vec{\theta})}$ depends on the particular instance of the algorithm. For quantum chemistry, the unitary coupled cluster (UCC) ansatz family is a popular choice \cite{anandQuantumComputingView2022}.

For the electronic structure problem, the Hamiltonian and/or ansatz operators are typically represented in the second quantization formalism, i.e., in terms of the fermionic creation/annihilation operators $a_i^\dagger$, $a_i$. A fermion-to-qubit mapping allows measuring/implementing these operators on quantum hardware, with a popular choice being the Jordan-Wigner transform \cite{jordanUeberPaulischeAequivalenzverbot1928}:

\begin{align}
\begin{split}
a_i^\dagger\rightarrow \frac{1}{2}\prod_{k=1}^{i-1}Z_k\cdot(X_i-iY_i),\\
\quad
a_i\rightarrow \frac{1}{2}\prod_{k=1}^{i-1}Z_k\cdot(X_i+iY_i),
\end{split}
\label{def:jw_transform_paulis}
\end{align}
where $Z_k$, $X_i$, and $Y_i$ are Pauli operators acting on the qubits labeled by the respective indices. We will use the Jordan-Wigner transform for all operators throughout this manuscript.

\subsection{ADAPT-VQE}
\label{ss:adaptvqe}

\begin{figure}[htbp]
    \centering
    \includegraphics[width=\columnwidth]{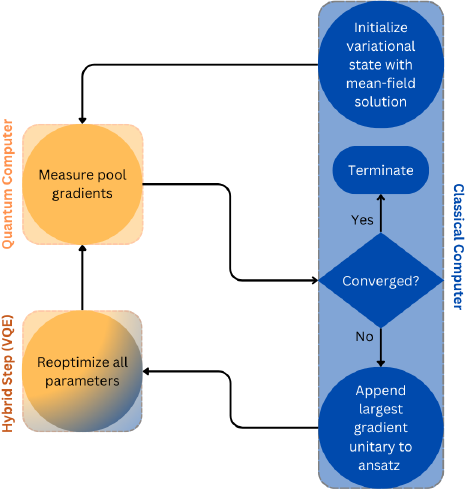}
    \caption{Flowchart for the ADAPT-VQE algorithm.}
    \label{fig:adapt}
\end{figure}

ADAPT-VQE, first proposed in Ref.~\cite{grimsleyAdaptiveVariationalAlgorithm2019}, is a special instance of the VQE algorithm where the ansatz is created dynamically instead of being defined \textit{a priori}. This protocol has been shown to outperform static VQE variants in circuit depth, measurement costs, and trainability \cite{grimsleyAdaptiveVariationalAlgorithm2019,ramoaReducingResourcesRequired2024}. We represent the protocol in Fig.~\ref{fig:adapt}.

The algorithm starts with a simple reference state. This is typically the Hartree-Fock ground state obtained from classical mean-field calculations, which corresponds to a computational basis state. Unitary operators are then added to the ansatz iteratively. These operators are of the form $e^{\hat{A}_k\theta_k}$, where the $\hat{A}_k$ are anti-hermitian generators belonging to a pre-selected operator pool. 

To decide which operator is added in each iteration, a gradient-based selection criterion is used. More specifically, the candidate operator with the energy derivative of highest magnitude at point zero is appended to the ansatz in each iteration. At iteration $n$, the gradient associated with a generator $\hat{A}_k$ can be evaluated as
\begin{equation}
    \pdv{E^{(n)}}{\theta_k}\Bigg|_{\theta_k=0} = \bra{\psi^{(n-1)}}\left[\mathcal{H},\hat{A}_k\right]\ket{\psi^{(n-1)}},
    \label{eq:commutator}
\end{equation}
where $\ket{\psi^{(n-1)}}$ is the final state from the previous iteration, and $\mathcal{H}$ is the Hamiltonian of the target system. The gradient criterion reflects the expectation that operators with a high impact on the energy locally will also have a high impact once a full optimization is performed. After each addition, the ansatz is fully optimized; the optimized parameters carry over as initial parameters for the next iteration. The new parameter is initialized at zero, which results in trainability benefits \cite{grimsleyADAPTVQEInsensitiveRough2022a} and allows for strategies such as Hessian recycling \cite{ramoaReducingMeasurementCosts2024b}. The algorithm ends when the norm of the vector formed from all gradients falls below a given threshold; throughout this work, we use the L2-norm to determine convergence. 

The operator pool is an important component of ADAPT-VQE, since it has an impact on the final circuit. The first proposal was to use a pool of fermionic excitation operators. An example single fermionic excitation is
\begin{equation}
\begin{split}
T_{\alpha_1\rightarrow\alpha_2}^{(FE)} = 
&a^\dagger_{\alpha_2}a_{\alpha_1} - a^\dagger_{\alpha_1}a_{\alpha_2}\\
\rightarrow&\frac{i}{2}\bigotimes_{z_k\in \mathcal{Z}_{JW}} Z_{z_k}\\
(&+X_{\alpha_2}Y_{\alpha_1}\\
&-Y_{\alpha_2}X_{\alpha_1}),\\
    \label{eq:fe_single}
\end{split}
\end{equation}
while an example fermionic double excitation is
\begin{equation}
\begin{split}
T_{\alpha_1\beta_1\rightarrow\alpha_2\beta_2}^{(FE)} = 
&a^\dagger_{\alpha_2}a^\dagger_{\beta_2}a_{\alpha_1}a_{\beta_1} - a^\dagger_{\beta_1}a^\dagger_{\alpha_1}a_{\beta_2}a_{\alpha_2}\\
\rightarrow&\frac{i}{8}\bigotimes_{z_k\in \mathcal{Z}_{JW}} Z_{z_k}\\
(&+X_{\alpha_2}X_{\alpha_1}X_{\beta_2}Y_{\beta_1}\\
&-X_{\alpha_2}X_{\alpha_1}Y_{\beta_2}X_{\beta_1}\\
&+X_{\alpha_2}Y_{\alpha_1}X_{\beta_2}X_{\beta_1}\\
&+X_{\alpha_2}Y_{\alpha_1}Y_{\beta_2}Y_{\beta_1}\\
&-Y_{\alpha_2}X_{\alpha_1}X_{\beta_2}X_{\beta_1}\\
&-Y_{\alpha_2}X_{\alpha_1}Y_{\beta_2}Y_{\beta_1}\\
&+Y_{\alpha_2}Y_{\alpha_1}X_{\beta_2}Y_{\beta_1}\\
&-Y_{\alpha_2}Y_{\alpha_1}Y_{\beta_2}X_{\beta_1}),\\
    \label{eq:fe_double}
\end{split}
\end{equation}
where $\mathcal{Z}_{JW}$ is the set of indices of qubits involved in the Jordan-Wigner anticommutation string arising from the Pauli $Z$ operators in Eq.~\eqref{def:jw_transform_paulis}, which depends on the qubit ordering chosen for the transform. Taking $i, j \leftarrow \texttt{sorted}(\alpha_1,\alpha_2)$ and $p, q, r, s \leftarrow \texttt{sorted}(\alpha_1,\alpha_2,\beta_1,\beta_2)$, we have $\mathcal{Z}_{JW} = \{x\in \mathbb{N}: i<x<j\}$ and $\mathcal{Z}_{JW} = \{x\in \mathbb{N}: p<x<q \vee r<x<s\}$ for the single and double excitations, respectively. We call the set of fermionic excitations of the form of Eqs.~\eqref{eq:fe_single} and \eqref{eq:fe_double} without index restrictions the generalized singles and doubles (GSD) pool.

The chain of Pauli $Z$ operators that arises when we apply the Jordan-Wigner transform to a fermionic excitation acts on a number of qubits that grows linearly on average with the size of the system. A consequence of this is that the unitaries generated by operators such as the ones in Eqs.~\eqref{eq:fe_single}, \eqref{eq:fe_double} require a linear number of entangling gates to be implemented. In order to improve the near-term viability of ADAPT-VQE, a pool of excitations without the anticommutation string has been proposed: the Qubit Excitation (QE) pool \cite{yordanovQubitexcitationbasedAdaptiveVariational2021}. The operators in this pool are formulated in terms of the qubit ladder operators:
\begin{align}
\begin{split}
&Q_i^\dagger = \frac{1}{2}(X_i-iY_i),\\
&Q_i = \frac{1}{2}(X_i+iY_i),
    \label{def:qubit_ladder_operators}
\end{split}
\end{align}
which are equivalent to the Jordan-Wigner-transformed fermionic ladder operators (Eq.~\eqref{def:jw_transform_paulis}) up to the anticommutation string. With these operators, we can define a qubit single excitation as
\begin{equation}
\begin{split}
T_{\alpha_1\rightarrow\alpha_2}^{(QE)} = 
&Q^\dagger_{\alpha_2}Q_{\alpha_1} - Q^\dagger_{\alpha_1}Q_{\alpha_2}\\
=\frac{i}{2}(+&X_{\alpha_2}Y_{\alpha_1}\\
&-Y_{\alpha_2}X_{\alpha_1}),\\
    \label{eq:qe_single}
\end{split}
\end{equation}
and a qubit double excitation as
\begin{equation}
\begin{split}
T_{\alpha_1\beta_1\rightarrow\alpha_2\beta_2}^{(QE)} = 
&Q^\dagger_{\alpha_2}Q^\dagger_{\beta_2}Q_{\alpha_1}Q_{\beta_1} - Q^\dagger_{\beta_1}Q^\dagger_{\alpha_1}Q_{\beta_2}Q_{\alpha_2}\\
=\frac{i}{8}(&+X_{\alpha_2}X_{\alpha_1}X_{\beta_2}Y_{\beta_1}\\
&-X_{\alpha_2}X_{\alpha_1}Y_{\beta_2}X_{\beta_1}\\
&+X_{\alpha_2}Y_{\alpha_1}X_{\beta_2}X_{\beta_1}\\
&+X_{\alpha_2}Y_{\alpha_1}Y_{\beta_2}Y_{\beta_1}\\
&-Y_{\alpha_2}X_{\alpha_1}X_{\beta_2}X_{\beta_1}\\
&-Y_{\alpha_2}X_{\alpha_1}Y_{\beta_2}Y_{\beta_1}\\
&+Y_{\alpha_2}Y_{\alpha_1}X_{\beta_2}Y_{\beta_1}\\
&-Y_{\alpha_2}Y_{\alpha_1}Y_{\beta_2}X_{\beta_1}).\\
    \label{eq:qe_double}
\end{split}
\end{equation}
Note that the operators in Eqs.~\eqref{eq:qe_single} and \eqref{eq:qe_double} are equivalent to those in Eqs.~\ref{cir:fe_single} and \ref{cir:fe_double} with the Pauli $Z$ operators removed.

The QEB-ADAPT-VQE protocol, which uses these operators instead of their fermionic versions, has been shown to lead to a reduction of the \texttt{CNOT} count with respect to the original fermionic pool \cite{yordanovQubitexcitationbasedAdaptiveVariational2021}. 

Recently, Ref.~\cite{ramoaReducingResourcesRequired2024} proposed a new variant of the algorithm, CEO-ADAPT-VQE, and showed that it further reduces the \texttt{CNOT} count with respect to QEB-ADAPT-VQE. This algorithm is based on coupled exchange operators (CEOs), which correspond to linear combinations of QEs with particularly efficient circuit implementations. CEO-ADAPT-VQE is the most hardware-efficient variant of ADAPT-VQE to date.

\subsection{Excitation Circuits}
\label{ss:circuits}

The unitaries generated by fermionic and qubit excitations allow for more efficient circuit implementations than unitaries generated by general linear combinations of Pauli strings \cite{yordanovEfficientQuantumCircuits2020a}. Throughout this work, we will use $U$ and $T$ to denote unitaries and the corresponding generators, respectively, such that, e.g., $U_{\alpha_1\rightarrow\alpha_2}^{(QE)}(\theta) = e^ {\theta  T_{\alpha_1\rightarrow\alpha_2}^{(QE)}}$.

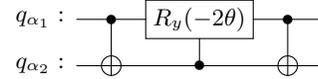
\begin{figure}[htbp]

    \centerline{
    \Qcircuit @C=1em @R=.7em {
    \lstick{q_{{\alpha}_{1}} :  } & \ctrl{1} &  \gate{R_y(-2\theta)} & \ctrl{1} & \qw \\
    \lstick{q_{{\alpha}_{2}} :  }  & \targ & \ctrl{-1} & \targ & \qw \\
    }}
    
    \caption{Circuit implementation of the unitary $U_{\alpha_1\rightarrow\alpha_2}^{(QE)}(\theta)$, generated by qubit excitation $T_{\alpha_1\rightarrow\alpha_2}^{(QE)}$. The same circuit can be used for $U_{\beta_1\rightarrow\beta_2}^{(QE)}$. The circuit for $U_{\alpha_2\rightarrow\alpha_1}^{(QE)}$ (or $U_{\beta_2\rightarrow\beta_1}^{(QE)}$) is identical, but with the rotation angle flipped.}

\label{cir:qe_single}
\end{figure}

The most efficient circuit implementation for the qubit single excitation defined in Eq.~\eqref{eq:qe_single} is shown in Fig.~\ref{cir:qe_single}. This circuit implements a $Y$ rotation if the states of the qubits representing the spin-orbitals involved in the excitation have odd parity, corresponding to an occupied/unoccupied spin-orbital pair where an excitation is allowed to occur. The controlled rotation gate can be implemented using two \texttt{CNOT}s and single-qubit gates; however, it is possible to optimize the final circuit such that it requires a total of only two \texttt{CNOT}s .

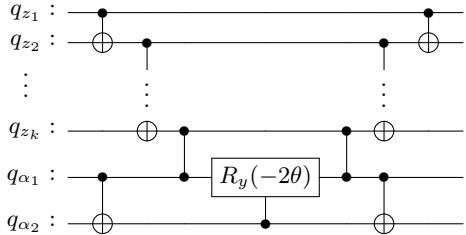
\begin{figure}[htbp]

    \centerline{
    \Qcircuit @C=1em @R=.7em {
    \lstick{q_{{z}_{1}} :  } & \ctrl{1} & \qw & \qw &  \qw & \qw & \qw & \ctrl{1} & \qw \\
    \lstick{q_{{z}_{2}} :  } & \targ & \ctrl{1} & \qw &  \qw & \qw & \ctrl{1} & \targ & \qw \\
     &&&&&&\\
      \lstick{\raisebox{7pt}{\vdots\hspace{10pt}}} & & \vdots & & & & \vdots \\
    \\
    \lstick{q_{{z}_{k}} :  } & \qw & \targ & \ctrl{1} & \qw & \ctrl{1} & \targ & \qw & \qw \\
    \lstick{q_{{\alpha}_{1}} :  } & \ctrl{1} & \qw & \ctrl{0} &  \gate{R_y(-2\theta)} & \ctrl{0} & \ctrl{1} & \qw & \qw \\
    \lstick{q_{{\alpha}_{2}} :  }  & \targ & \qw & \qw & \ctrl{-1} & \qw & \targ & \qw & \qw \\
    }}
    
    \caption{Circuit implementation of the unitary $U_{\alpha_1\rightarrow\alpha_2}^{(FE)}(\theta)$, generated by qubit excitation $T_{\alpha_1\rightarrow\alpha_2}^{(FE)}$. The same circuit can be used for $U_{\beta_1\rightarrow\beta_2}^{(FE)}$. The circuit for $U_{\alpha_2\rightarrow\alpha_1}^{(FE)}$ (or $U_{\beta_2\rightarrow\beta_1}^{(FE)}$) is identical, but with the rotation angle flipped. The vertical dots indicate that there is a ladder of \texttt{CNOT} gates between qubits $q_{{z}_1}$ to $q_{|{\cal Z}_{JW}|}$.}

\label{cir:fe_single}
\end{figure}

In contrast, the circuit required to implement the single fermionic excitation in Eq.~\eqref{eq:fe_single} is depicted in Fig.~\ref{cir:fe_single}. As compared to the qubit single excitation circuit, we require additional \texttt{CNOT} gates to calculate the parity of the set $\mathcal{Z}_{JW}$ of qubits involved in the Jordan-Wigner anticommutation string, as discussed in the previous section. The number of \texttt{CNOT}s required to implement this circuit is $\mathcal{O}(N)$, where $N$ is the total number of qubits. The specific \texttt{CNOT} count depends on how many qubits appear between the qubits $q_{\alpha_1}$ and $q_{\alpha_2}$ in the ordering considered for the Jordan-Wigner transform.

\begin{figure}[htbp]

    \centerline{
    \Qcircuit @C=1em @R=.7em {
    \lstick{q_{{\beta}_{1}} :  } & \ctrl{2} & \qw & \ctrl{1} & \gate{R_y(-2\theta)} & \ctrl{1} & \qw & \ctrl{2} & \qw \\
    \lstick{q_{{\beta}_{2}} :  } & \qw & \ctrl{2} & \targ & \ctrl{-1}& \targ  & \ctrl{2} & \qw & \qw  \\
    \lstick{q_{{\alpha}_{1}} :  } & \targ & \qw & \qw & \ctrlo{-1} & \qw & \qw & \targ  & \qw &\\
    \lstick{q_{{\alpha}_{2}} :  } & \qw & \targ & \qw & \ctrlo{-1}& \qw  & \targ & \qw & \qw
    }}
    
    \caption{Circuit implementation of the unitary $U_{\alpha_1\beta_1\rightarrow\alpha_2\beta_2}^{(QE)}(\theta)$, generated by qubit excitation $T_{\alpha_1\beta_1\rightarrow\alpha_2\beta_2}^{(QE)}$. The circuit for $U_{\alpha_2\beta_2\rightarrow\alpha_1\beta_1}^{(QE)}$ is identical, but with the rotation angle flipped.}

\label{cir:qe_double}
\end{figure}
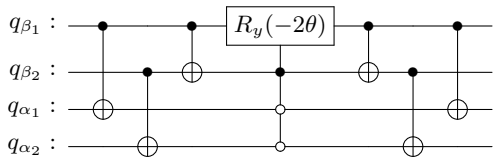

While single excitations can be implemented using one-control $Y$ rotations, double excitations require three-control $Y$ rotations. The circuit implementation of the qubit double excitation in Eq.~\eqref{eq:qe_double} is shown in Fig.~\ref{cir:qe_double}. \texttt{CNOT}s are used to calculate the parities of the states of subsets (pairs) of qubits, such that the controlled rotation only occurs if spin-orbitals $\alpha_1, \beta_1$ are occupied and $\alpha_2, \beta_2$ are unoccupied, or vice-versa. Once the multi-control rotation is decomposed into basic gates and circuit optimizations are applied, the circuit includes a total of 13 \texttt{CNOT}s. 

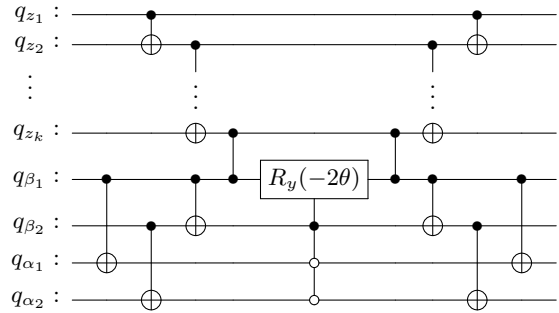
\begin{figure}[htbp]

    \centerline{
    \Qcircuit @C=1em @R=.7em {
    \lstick{q_{{z}_{1}} :  } & \qw & \ctrl{1} & \qw & \qw &  \qw & \qw & \qw & \ctrl{1} & \qw & \qw \\
    \lstick{q_{{z}_{2}} :  } & \qw & \targ & \ctrl{1} & \qw &  \qw & \qw & \ctrl{1} & \targ & \qw & \qw \\
     &&&&&&&\\
      \lstick{\raisebox{7pt}{\vdots\hspace{10pt}}} & & &  \vdots & & & & \vdots \\
    \\
    \lstick{q_{{z}_{k}} :  } & \qw & \qw & \targ & \ctrl{1} & \qw &  \ctrl{1} & \targ & \qw & \qw & \qw \\
    \lstick{q_{{\beta}_{1}} :  } & \ctrl{2} & \qw & \ctrl{1} & \ctrl{0} & \gate{R_y(-2\theta)} & \ctrl{0} & \ctrl{1} & \qw & \ctrl{2} & \qw \\
    \lstick{q_{{\beta}_{2}} :  } & \qw & \ctrl{2} & \targ & \qw & \ctrl{-1} & \qw & \targ & \ctrl{2} & \qw & \qw  \\
    \lstick{q_{{\alpha}_{1}} :  } & \targ & \qw & \qw & \qw & \ctrlo{-1} & \qw & \qw & \qw & \targ  & \qw &\\
    \lstick{q_{{\alpha}_{2}} :  } & \qw & \targ & \qw & \qw & \ctrlo{-1} & \qw & \qw  & \targ & \qw & \qw
    }}
    
    \caption{Circuit implementation of the unitary $U_{\alpha_1\beta_1\rightarrow\alpha_2\beta_2}^{(FE)}(\theta)$, generated by qubit excitation $T_{\alpha_1\beta_1\rightarrow\alpha_2\beta_2}^{(FE)}$. The circuit for $U_{\alpha_2\beta_2\rightarrow\alpha_1\beta_1}^{(FE)}$ is identical, but with the rotation angle flipped. The vertical dots indicate that there is a ladder of \texttt{CNOT} gates between qubits $q_{{z}_1}$ to $q_{|{\cal Z}_{JW}|}$.}

\label{cir:fe_double}
\end{figure}

Similar to the case of single excitations, the fermionic double excitation in Eq.~\eqref{eq:fe_double} can be implemented using the qubit double excitation circuit with additional \texttt{CNOT}s to calculate the parity of the states of qubits involved in the Jordan-Wigner anticommutation string. The resulting circuit is depicted in Fig.~\ref{cir:fe_double}. Once again, we have a $\mathcal{O}(N)$ circuit count, with the specific number depending on the spin-orbital ordering.

CEOs consist of linear combinations of QEs that allow for particularly efficient circuit implementations \cite{ramoaReducingResourcesRequired2024}. They come in two flavors: OVP-CEOs, which consist of the sum or the difference of two double qubit excitations sharing a variational parameter, and MVP-CEOs, which consist of linear combinations of up to three double qubit excitations with independent variational parameters.

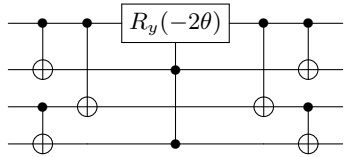
\begin{figure*}[htbp]

    \centerline{
    \Qcircuit @C=1em @R=.7em {
    & \ctrl{1} & \ctrl{2} & \gate{R_y(-2\theta)} & \ctrl{2} & \ctrl{1} & \qw \\
    & \targ & \qw & \ctrl{-1}& \qw & \targ & \qw  \\
    & \ctrl{1} & \targ & \qw & \targ & \ctrl{1} & \qw\\
    & \targ & \qw & \ctrl{-2}& \qw & \targ  & \qw 
    }}
    
    \caption{Circuit Implementation of $U_{\alpha_1\beta_1\alpha_2\beta_2}^{(OVP-CEO,+)}(\theta)$, a unitary generated by a coupled exchange operator consisting of the sum of two qubit excitations.}

\label{cir:ceo_implicit}
\end{figure*}

\begin{figure*}[htbp]

    \footnotesize
    \centerline{
    \Qcircuit @C=0.3em @R=.7em {
    &\gate{S^\dagger} & \ctrl{3} & \ctrl{2} & \ctrl{1} & \gate{H} & \gate{R_z\left(-\theta_0\right)} & \targ & \gate{R_z\left(-\theta_1\right)} & \targ & \gate{R_z\left(\theta_2\right)} & \targ & \gate{R_z\left(-\theta_3\right)} & \targ & \gate{R_z\left(\theta_4\right)} & \targ & \gate{R_z\left(\theta_5\right)} & \targ & \gate{R_z\left(\theta_6\right)} & \targ & \gate{R_z\left(-\theta_7\right)} & \gate{H} & \ctrl{1} & \ctrl{2} & \ctrl{3} & \qw & \qw  \\
    & \qw & \qw & \qw & \targ & \qw & \qw & \ctrl{-1} & \qw & \qw & \qw & \ctrl{-1} & \qw & \qw & \qw & \ctrl{-1} & \qw & \qw & \qw & \ctrl{-1} & \qw & \qw & \targ & \qw & \qw  & \qw  & \qw \\
    & \qw & \qw & \targ & \qw &\gate{S^\dagger} & \qw & \qw & \qw & \qw & \qw & \qw & \qw & \ctrl{-2} & \qw & \qw & \qw & \qw & \qw & \qw & \qw & \qw & \qw & \targ & \qw &\gate{S} & \qw \\
    & \qw & \targ & \qw & \qw & \qw & \qw & \qw & \qw & \ctrl{-3} & \qw & \qw & \qw & \qw & \qw & \qw & \qw &\ctrl{-3} \qw & \qw & \qw 
    & \qw & \qw & \qw & \qw & \targ & \qw  & \qw }}
    
    \caption{Circuit implementation of $e^{\frac{i}{8} (\theta_0XXXY + \theta_1XXYX + \theta_2YXYY + \theta_3YXXX + \theta_4YYXY + \theta_5YYYX + \theta_6XYYY + \theta_7XYXX)}$. Unitaries generated by a linear combination of the eight Pauli strings that appear in a double qubit excitation take this form.}

\label{cir:ceo_ind_params}
\end{figure*}
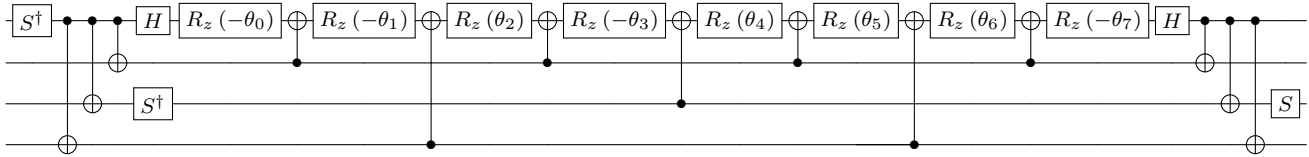

For example, the operator
\begin{align}
\begin{split}
&T_{\alpha_1\beta_1\alpha_2\beta_2}^{(OVP-CEO,+)}(\theta)=
\theta(T_{\alpha_1\beta_1\rightarrow\alpha_2\beta_2}^{(QE)} + T_{\alpha_2\beta_1\rightarrow\alpha_1\beta_2}^{(QE)}) \\
&= \theta(Q^\dagger_{\alpha_2}Q^\dagger_{\beta_2}Q_{\alpha_1}Q_{\beta_1} + Q^\dagger_{\alpha_1}Q^\dagger_{\beta_2}Q_{\alpha_2}Q_{\beta_1}) - h.c.
\end{split}
\label{eq:ceo_sum}
\end{align}
generates a unitary which can be implemented with the circuit in Fig.~\ref{cir:ceo_implicit}. This circuit can be implemented using 9 \texttt{CNOT}s and single-qubit gates, offering a 30\% decrease in two-qubit gate count as compared to a unitary generated by one of the constituent qubit excitations.
An even greater reduction can be achieved in the case of MVP-CEOs. These operators may include up to three independently parameterized double qubit excitations. As long as these excitations act on the same set of four spin-orbitals, the resulting generator is of the form $\theta_0XXXY + \theta_1XXYX + \theta_2YXYY + \theta_3YXXX + \theta_4YYXY + \theta_5YYYX + \theta_6XYYY + \theta_7XYXX$ (where the parameters $\theta_0$,...,$\theta_7$ depend on up to three independent parameters) and can be implemented with the 13-CNOT circuit in Fig.~\ref{cir:ceo_ind_params}, which translates to a 67\% improvement with respect to the 39 \texttt{CNOT}s required to implement the three unitaries generated by three generic double qubit excitations.

\section{Co-ADAPT-VQE}
\label{s:co-adapt}

The canonical ADAPT-VQE protocol (Fig.~\ref{fig:adapt}) appends to the ansatz the operator with the highest gradient in each iteration. This is an attempt to maximize the impact on the energy per operator. However, such a selection criterion is architecture-agnostic; it does not take into account important factors such as how hard it is to implement each operator in the target hardware, or how affected by noise each operator is. Operators that require interactions between distant physical qubits (in the target hardware's connectivity graph), or that act on qubits with higher noise rates and lower coherence times, may be poor choices even if the gradients are high; in some cases, the impact on the energy may not compensate the cost of implementing this operator in terms of gate count, additional impact of noise, increase in circuit depth, or other factors.

As such, we propose to modify the ADAPT-VQE selection criterion to include information about the quantum computer on which the algorithm is meant to be executed. Specifically, we propose to scale the gradients as 
\begin{equation}
    \frac{\pdv{E^{(n)}}{\theta_i}\Bigg|_{\theta_i=0}}{p} = \frac{\bra{\psi^{(n-1)}}\left[\hat{H},\hat{A}_k\right]\ket{\psi^{(n-1)}}}{p},
    \label{eq:penalty}
\end{equation}
where \textit{p} is a hardware-dependent penalty. This penalty may be related to any relevant metric for the cost of implementing the operator in the target quantum computer, e.g., gate counts after transpilation into the target gate set and connectivity graph, average coherence times of the qubits involved in the operation, or summed average errors for all two-qubit interactions, etc. In these examples, the penalty for each operator may be calculated before the start of the protocol and stored for use in each ADAPT-VQE iteration; however, the penalty may also be recalculated dynamically as the algorithm evolves. An example is if we wish to minimize circuit depth; in this case, we may recalculate the penalty in each iteration such that it corresponds to the additional circuit depth we get when appending each operator to the current ansatz circuit. Note, however, that while polynomial, the classical overhead stemming from this dynamic calculation may be significant if we perform circuit optimization after the attempted addition of each candidate operator, since the total number of circuit optimizations per iteration will then be equal to the number of operators in the pool, which scales as $\mathcal{O}(N^4)$ for the best performing pools, where $N$ is the number of orbitals (or qubits) \cite{ramoaReducingResourcesRequired2024,yordanovQubitexcitationbasedAdaptiveVariational2021}.

\section{Using Co-ADAPT-VQE to Minimize Entangling Gate Counts for LNN Connectivity}
\label{s:minimize_lnn_cnots}

We will exemplify the impact of our co-designed variant of ADAPT-VQE (Co-ADAPT-VQE) by using it to minimize the entangling gate count of the state preparation circuits run on hardware with LNN connectivity. This task has broad relevance: Since entangling gates are associated with the highest error rates, their number is a good indicator of how a circuit will be impacted by hardware noise \cite{daltonQuantifyingEffectGate2024}. Further, linear connectivity is a common target: for example, the connectivity graph for superconducting architectures is typically close to a linear graph, with the average number of connections per qubit (degree of the connectivity graph) very close to 2 \cite{abughanemIBMQuantumComputers2025}. Further, while neutral atom and trapped-ion architectures offer ATA connectivity, in practice these interactions rely on the spatial rearrangement of qubits, such that significant overheads are incurred in shuttling/transport operations \cite{delaneyScalableMultispeciesIon2024,pinoDemonstrationTrappedionQuantum2021,hensingerQuantumComputerBased2021,decrossComputationalPowerRandom2025}.

\begin{figure*}
    \centering
    \includegraphics[width=0.8\linewidth]{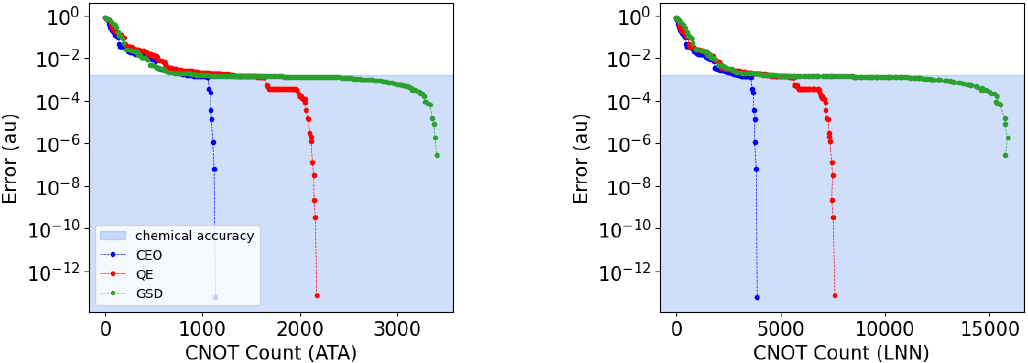}
    \caption{Error \textit{versus} \texttt{CNOT} count for the best performing variants of ADAPT-VQE: CEO-, QE- and Qubit-ADAPT-VQE. The molecule considered is H$_6$ at interatomic distance 3 \AA{} in the STO-3G basis set. The blue shading represents the chemical accuracy region (error below 1 kcal/mol).}
    \label{fig:naive_transpile}
\end{figure*}

In Sec.~\ref{s:background}, ATA connectivity was assumed for all circuits. This has been the standard for benchmarking ADAPT-VQE variants \cite{tangQubitADAPTVQEAdaptiveAlgorithm2021,yordanovQubitexcitationbasedAdaptiveVariational2021,ramoaReducingResourcesRequired2024}. Transpilation to LNN connectivity and circuit optimization can be straightforwardly applied to any ansatz. Fig.~\ref{fig:naive_transpile} shows that the relative performance of the pools remains the same upon transpilation of the original ATA circuits, with CEO-ADAPT-VQE \cite{ramoaReducingResourcesRequired2024} resulting in the lowest two-qubit gate count, followed by QE-ADAPT-VQE \cite{yordanovQubitexcitationbasedAdaptiveVariational2021} and finally Qubit-ADAPT-VQE \cite{tangQubitADAPTVQEAdaptiveAlgorithm2021}.

While the relative merits of the algorithmic variants remain similar upon transpilation from ATA to LNN connectivity, Fig.~\ref{fig:naive_transpile} tells us that transpilation incurs a roughly 4-fold increase in the \texttt{CNOT} count---an increase which is expected to aggravate with the system size. We will showcase the impact of Co-ADAPT-VQE by using it to mitigate this overhead. We will focus on transpilation to a gate set consisting of single-qubit rotations and \texttt{CNOT}s for LNN connectivity; however, the same techniques may be readily applied to other hardware specifications.

Typical routing protocols are based on networks of \texttt{SWAP} gates, which are two-qubit gates with matrix representation

\[
\texttt{SWAP} \equiv
\begin{bmatrix}
1 & 0 & 0 & 0 \\
0 & 0 & 1 & 0 \\
0 & 1 & 0 & 0 \\
0 & 0 & 0 & 1
\end{bmatrix}.
\]

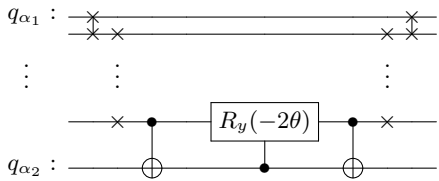
\begin{figure}[htbp]

    \centerline{
    \Qcircuit @C=1em @R=.7em {
    \lstick{q_{{\alpha}_{1}} :  } & \qswap & \qw & \qw & \qw &  \qw & \qw & \qw & \qswap & \qw \\
     & \qswap \qwx[-1] & \qswap & \qw & \qw & \qw &  \qw & \qswap & \qswap\qwx[-1] & \qw \\
     &&&&&&\\
      \lstick{{\vdots\hspace{10pt}}} & & \vdots & & & & & \vdots \\
    \\
    & \qw & \qswap & \ctrl{1} & \qw &  \gate{R_y(-2\theta)} & \ctrl{1} & \qswap & \qw & \qw \\
    \lstick{q_{{\alpha}_{2}} :  } & \qw & \qw & \targ & \qw & \ctrl{-1} & \targ & \qw & \qw & \qw \\
    }}
    
    \caption{Implementation of the qubit single excitation circuit in Fig.~\ref{cir:qe_single} for LNN connectivity using swap-based routing. Unmarked qubits are irrelevant to the excitation, but they are part of the path from the original to the final physical qubit representing qubit $\alpha_1$.}

\label{cir:qe_single_lnn}
\end{figure}

A \texttt{SWAP} gate acting on qubits $i$, $j$ exchanges $\ket{0}_i\ket{1}_j\leftrightarrow \ket{1}_i\ket{0}_j$, allowing us to modify the qubit ordering by exchanging the states of qubit pairs. As a specific example, let us consider the qubit single-excitation circuit from Fig.~\eqref{eq:qe_single}. If qubits $q_{\alpha_1}$ and $q_{\alpha_2}$ are not adjacent in the connectivity graph, then we cannot apply the \texttt{CNOT} gates directly. However, we can route the qubits together using a ladder of \texttt{SWAP} gates, as depicted in Fig.~\ref{cir:qe_single_lnn}. After we apply the desired operations, the adjoint \texttt{SWAP} ladder is applied to restore the original ordering such that the unitaries exactly match.

\begin{figure}[htbp]
\centerline{
\Qcircuit @C=1.0em @R=0.8em @!R { \\
	 	& \qswap & \qw & & & \ctrl{1} & \targ & \ctrl{1} & \qw\\
	 	& \qswap \qwx[-1] & \qw & \raisebox{1.5em}{=} & & \targ & \ctrl{-1} & \targ & \qw\\
\\ }
}
\caption{Decomposition of a \texttt{SWAP} gate into \texttt{CNOT} gates.}
\label{cir:swap}
\end{figure}
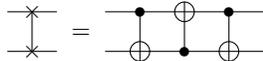

Note that \texttt{SWAP} gates are not part of the gate set we are considering; however, they can be implemented using \texttt{CNOT} gates as shown in Fig.~\ref{cir:swap}.

Qubit double excitations can be transpiled similarly; particularly, for a set of sorted physical qubits $[p,q,r,s]$, we take the central point $k=\texttt{round}(\frac{p + q + r + s}{2})$ (where \texttt{round}$(i)$ returns the nearest integer to $i$) and route them to $[k-1,k,k+1,k+2]$ or $[k-2,k-1,k,k+1]$ if the fraction was rounded up or down, respectively. \texttt{CNOT} gates within the 4-qubit circuit can be straightforwardly transpiled using similar methods. 

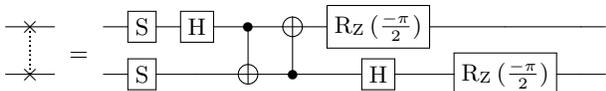
\begin{figure}[htbp]
\centerline{
\Qcircuit @C=1.0em @R=0.2em @!R { \\
	 	& \qswap & \qw & & & \gate{\mathrm{S}} & \gate{\mathrm{H}} & \ctrl{1} & \targ & \gate{\mathrm{R_Z}\,(\mathrm{\frac{-\pi}{2}})} & \qw & \qw & \qw\\
	 	& \qswap \qwxdotted[-1] & \qw & \raisebox{1.5em}{=} & & \gate{\mathrm{S}} & \qw & \targ & \ctrl{-1} & \gate{\mathrm{H}} & \gate{\mathrm{R_Z}\,(\mathrm{\frac{-\pi}{2}})} & \qw & \qw\\
\\ }
}
\caption{Decomposition of an \texttt{FSWAP} gate into \texttt{CNOT} gates.}
\label{cir:fswap}
\end{figure}

Naively, we could apply the same \texttt{SWAP}-based routing method for fermionic excitations. However, the \texttt{FSWAP} gate proposed in Ref.~\cite{kivlichanQuantumSimulationElectronic2018} enables a more efficient transpilation protocol, specifically tailored to fermionic operators. This gate is defined as

\[
\texttt{FSWAP} \equiv
\begin{bmatrix}
1 & 0 & 0 & 0 \\
0 & 0 & 1 & 0 \\
0 & 1 & 0 & 0 \\
0 & 0 & 0 & -1
\end{bmatrix},
\]
and corresponds to the exchange of \textit{fermionic modes} rather than qubits. The additional phase in the last entry of the matrix accounts for fermionic anticommutation effects. If fermionic modes $i$, $j$ had been mapped to qubits $a$, $b$ as $i\leftrightarrow a$ and $j\leftrightarrow b$, $\texttt{FSWAP}(a,b)$ reassigns the modes such that $i\leftrightarrow b$ and $j\leftrightarrow a$. Conveniently, this gate can be implemented using only 2 \texttt{CNOT} gates using the circuit in Fig.~\ref{cir:fswap}. We depict the \texttt{FSWAP} gate using a dashed line so that it can be visually distinguished from the \texttt{SWAP} gate.

\begin{figure}[htbp]

    \centerline{
    \Qcircuit @C=1em @R=.7em {
    \lstick{q_{{\alpha}_{1}} :  } & \qswap & \qw & \qw & \qw &  \qw & \qw & \qw & \qswap & \qw \\
    \lstick{q_{{z}_{1}} :  } & \qswap \qwxdotted[-1] & \qswap & \qw & \qw & \qw &  \qw & \qswap & \qswap\qwxdotted[-1] & \qw \\
     &&&&&&\\
      \lstick{\raisebox{7pt}{\vdots\hspace{10pt}}} & & \vdots & & & & & \vdots \\
    \\
    \lstick{q_{{z}_{k}} :  }& \qw & \qswap & \ctrl{1} & \qw &  \gate{R_y(-2\theta)} & \ctrl{1} & \qswap & \qw & \qw \\
    \lstick{q_{{\alpha}_{2}} :  } & \qw & \qw & \targ & \qw & \ctrl{-1} & \targ & \qw & \qw & \qw \\
    }}
    
    \caption{Implementation of the single fermionic excitation circuit in Fig.~\ref{cir:fe_single} for LNN connectivity. It is assumed that the mapping of fermionic modes to physical qubits is consistent with the ordering used in the Jordan-Wigner transform.}

\label{cir:fe_single_lnn}
\end{figure}
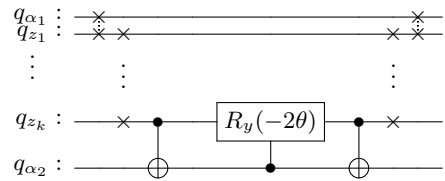

With this gate, we can implement the fermionic single excitation circuit of Fig.~\ref{cir:fe_single} using the circuit in Fig.~\ref{cir:fe_single_lnn}. Fermionic double excitations can be similarly implemented, as discussed above in the context of qubit double excitations.

While for ATA connectivity fermionic excitations require many more entangling gates ($\mathcal{O}(N)$, as opposed to $\mathcal{O}(1)$ for qubit excitations), for LNN connectivity they may actually require \textit{fewer} entangling gates than qubit excitations, due to the more efficient decomposition of the \texttt{FSWAP} gate.

\begin{table*}
\setlength{\tabcolsep}{0pt}
\begin{center}
\begin{tabular}{|c|c|c|c|c|c|c|c|c|c|c|}
\hline
 \rowcolor{gray1}\rotatebox[origin=c]{90}{ Implementation } & \multicolumn{5}{c|}{Product of Pauli Rotations} & \multicolumn{5}{c|}{Multi-Controlled Y Rotations}\\
 \hline
  \rowcolor{gray2} \rotatebox[origin=c]{90}{ Connectivity } & { All-to-All } & \multicolumn{4}{c|}{Linear Nearest Neighbor} & { All-to-All } & \multicolumn{4}{c|}{Linear Nearest Neighbor}\\
 \hline
  \rowcolor{gray3} \rotatebox[origin=c]{90}{ Routing } & - & \multicolumn{2}{c|}{Swaps} &  \multicolumn{2}{c|}{FSwaps} & - & \multicolumn{2}{c|}{Swaps} & \multicolumn{2}{c|}{FSwaps}\\
 \hline
  \rowcolor{gray4}\rotatebox[origin=c]{90}{ Final Layout } & - & { Fixed } &  { Variable } & { Fixed } &  { Variable } & - & { Fixed } &  { Variable } & { Fixed } &  { Variable }\\
 \hline
 \hiderowcolors\rotatebox[origin=c]{90}{ $a_0a_{99}^\dagger$ - h.c. } & 396 & 396 & 396 & 396 & 200 & 196 & 1020 & 729 & 394 & 198 \\ 
 \hline
 \rotatebox[origin=c]{90}{ $Q_0Q_{99}^\dagger$ - h.c. }  & 4 & 592 & 298 & - & - & 2 & 590 & 296 & - & - \\  
 \hline  
\end{tabular}
\end{center}
  \caption{Number of \texttt{CNOT} gates necessary to implement the unitaries generated by fermionic ($a_0a_{99}^\dagger$ - h.c.) and qubit ($Q_0Q_{99}^\dagger$ - h.c.) excitations. These unitaries act on a system with 100 spin-orbitals (equivalently, qubits), exchanging fermions from spin-orbitals 0 and 99. The table compares the two types of circuit implementations discussed in the main text under all-to-all and linear nearest neighbor connectivity. For the latter, two routing strategies are considered: one using \texttt{SWAP} gates and one using \texttt{FSWAP} gates (when applicable). Further, since it is possible to keep track of the qubit layout classically, we consider a setting where the final qubit layout is variable (i.e., equivalent to the initial one up to a permutation). All the circuits are transpiled using the respective techniques with no further optimization. While circuit optimization methods might produce slight improvements, they do not meaningfully change the trends observed in the table.}
  \label{tab:gates}
\end{table*}

Table \ref{tab:gates} showcases this difference by considering the \texttt{CNOT} counts necessary to implement a fermionic single excitation and its qubit counterpart on a 100-qubit system. Two distinct implementation strategies are considered: One based on multi-controlled $Y$ rotations followed by decomposition into a \texttt{CNOT} + single qubit rotations gate set, as proposed in Ref.~\cite{yordanovEfficientQuantumCircuits2020a} and discussed above, and one based on the general implementation of Pauli rotations using ladders of \texttt{CNOT} gates, which can be straightforwardly applied here since all Pauli strings in the generators commute \cite{Nielsen_Chuang}.

We observe that for ATA connectivity, qubit excitations are a by far the most gate-efficient option, requiring only 2--4 \texttt{CNOT}s as compared to the hundreds required by the fermionic counterparts. Yet, the number of gates required by qubit excitations increases significantly when the unitaries are transpiled to LNN connectivity. In particular, if we allow the final layout of qubits or fermionic modes to vary---allowing us to measure the same expectation values with shallower circuits by exchanging the final \texttt{SWAP}/\texttt{FSWAP} networks for classically tractable manipulation of the observables---we can implement the fermionic excitation with as few as 198 \texttt{CNOT} gates, while the qubit excitation requires at least 298.

The same strategy can be applied to other fermionic operator pools; in particular, we can create a fermionic version of CEOs by exchanging $Q_i^\dagger\rightarrow a_i^\dagger$, $Q_i\rightarrow a_i$ in their definition. These fermionic CEO operators can be readily implemented with the same \texttt{SWAP}-based circuits that would implement CEOs in a LNN architecture, by simply exchanging the \texttt{SWAP} gates for \texttt{FSWAP} gates. This technique applies to OVP- as well as MVP-CEOs.

With these ideas in mind, we design 3 variants of Co-ADAPT-VQE:

\begin{itemize}
    \item \texttt{SWAP}-based: This variant is suitable for local operators (i.e., operators acting on $\mathcal{O}(1)$ qubits). Examples of this are the qubit, QE and CEO pools, whose operators are either 2- or 4-local. The transpilation of each operator is based on a \texttt{SWAP} network which brings the relevant qubits together, followed by the implementation of the actual excitation circuit. After the excitation circuit, the physical qubit assignment is classically updated to avoid the application of an order-restoring \texttt{SWAP} network. The penalty applied to each operator according to Eq.~\eqref{eq:penalty} is the total \texttt{CNOT} count of the circuit when implemented as described. Since the qubit assignment is dynamic, the penalty associated with each operator is recalculated in each iteration. This penalty can be calculated classically without actually constructing the circuits: for a single excitation acting on qubits $a<b$, the total number of \texttt{SWAP} gates will be $b-a-1$, while for a double excitation acting on qubits $a<b<c<d$ it will be $|a-a'| + |b-b'| + |c-c'| + |d-d'|$, where $a', b', c', d'$ are the target physical qubits which can be obtained from $a, b, c, d$ as described earlier in this section. At the end of the circuit, the final qubit assignment is used to redefine the observable, which is equivalent to applying a final \texttt{SWAP} network while avoiding the associated gate overhead. This reassignment is performed by permuting qubit indices in the Pauli strings that appear in the observable.

    \item \texttt{FSWAP}-based: This variant is suitable for fermionic operators, such as the generalized singles and doubles pool or the fermionic version of the CEO pool described above. The implementation is similar to the \texttt{SWAP}-based variant, with the difference of using \texttt{FSWAP} networks instead of regular \texttt{SWAP} networks. This allows us to bring fermionic modes together and implement a fermionic excitation---including the anticommutation string---with fewer gates than would be required to implement the corresponding qubit excitation. Accordingly, what we update at the end of each iteration to avoid order-restoring \texttt{FSWAP} ladders is the assignment of fermionic modes to qubits. At the end of the circuit, the final mode-to-qubit assignment is used to redefine the observable. This reassignment is performed by permuting mode indices in the excitations that appear in the fermionic Hamiltonian, which is equivalent to applying a final \texttt{FSWAP} network while avoiding the associated gate overhead. Importantly, this step is done \textit{before} mapping the Hamiltonian under the Jordan-Wigner transform. The method used to calculate the penalty in this \texttt{FSWAP}-based variant is identical to the method used for the \texttt{SWAP}-based one, as described above. Since the mode assignment is dynamic, the penalty associated with each operator is recalculated in each iteration. As before, the penalty can be efficiently calculated classically without building the circuits.
    
    \item \texttt{Qiskit}-transpiler-based: This variant is fully generic, and it is suitable for any operator pool. In this case, the penalty associated with an operator is calculated by transpiling the corresponding circuit to the desired connectivity using \texttt{Qiskit} \cite{QiskitCommunity2017} under the maximum optimization level. In this case, the penalty is static: We can create the circuit associated with each pool operator at the start of the algorithm and store it along with the corresponding \texttt{CNOT} count so that it can be used for all iterations.
\end{itemize}

For all variants listed above, as well as for the regular ADAPT-VQE circuits transpiled to restricted connectivity, we apply the \texttt{Qiskit} \cite{QiskitCommunity2017} transpiler at maximum optimization level to the final circuit. As this transpiler changes the assignment of physical qubits, we update the final observable by exchanging the indices in the Pauli strings according to the permutation. In the case of the \texttt{SWAP}-based Co-ADAPT-VQE variant, this permutation straightforwardly composes with the one arising from the algorithm. In the case of the \texttt{FSWAP}-based Co-ADAPT-VQE variant, we apply the permutation of modes arising from the algorithm to the fermionic Hamiltonian, map the resulting operator to a linear combination of Pauli strings using the Jordan-Wigner transform, and finally permute the indices of the Pauli strings according to the physical qubit assignment output by the transpiler. In the case of the \texttt{Qiskit}-transpiler-based Co-ADAPT-VQE variant, as well as of the regular ADAPT-VQE variants transpiled to restricted connectivity, the only permutation is the one associated with the transpilation, which can be readily corrected for by remapping indices in the observable after the Jordan-Wigner transform.

Finally, we note that the routing strategies we discussed in this section go well beyond LNN connectivity, and connectivity graphs with a higher degree will allow for an even more efficient transpilation. As an example, let us consider the asymptotic scaling of the average distance between qubits in a linear graph (degree 2) as compared to a square lattice (degree 4). 

In the case of a linear graph of size $N$ ($N$ linearly connected qubits), we have a total of ${\binom{N}{2}}$ qubit pairs. For each distance $k\in\{1,..., N-1\}$ we have exactly $N-k$ pairs $(i,j)$ such that $j-i=k$. As such, the average distance between qubits is

\begin{equation}
\frac{1}{\binom{N}{2}} \sum_{k=1}^{N-1}k(N-k) = \frac{2}{N(N-1)}\frac{N^3-N}{6}
= \frac{N+1}{3}.
\end{equation}

In the case of a square lattice with $N$ qubits, the Manhattan distance $|x_1-x_2| + |y_1-y_2|$ gives us the shortest path between two qubits with coordinates $(x_1,y_1)$ and $(x_2,y_2)$. Since the two coordinates are independent, the total distance will be $2\frac{L+1}{3}$, where $L=\sqrt{N}$ is the maximum distance in either direction.

Hence, 2D connectivity offers an asymptotic improvement in the average distance between qubits (or equivalently, the average number of \texttt{SWAP} gates required for transpiling an operator) from $\mathcal{O}(N)$ to $\mathcal{O}(\sqrt{N})$, where $N$ is the number of qubits. Similar conclusions can be drawn for higher degrees of connectivity.

\section{Results}
\label{s:results}

This section presents simulation results confirming that the instances of Co-ADAPT-VQE defined above are effective at decreasing the \texttt{CNOT} count of state preparation circuits, as they were designed to do. For all systems, we use the minimal STO-3G basis set. All circuits are transpiled to a gate set consisting of single-qubit gates and \texttt{CNOT}s, and all simulation results are noise-free. We note that \texttt{CNOT} count was shown to be a good predictor of how much a given circuit will be impacted by noise \cite{daltonQuantifyingEffectGate2024}; hence, equating a lower \texttt{CNOT} count to more noise-resilient circuits is a standard assumption. The code used to conduct the simulations is publicly available at \url{https://github.com/mafaldaramoa/ceo-adapt-vqe}.

\begin{figure*}
    \centering
    \includegraphics[width=0.95\linewidth]{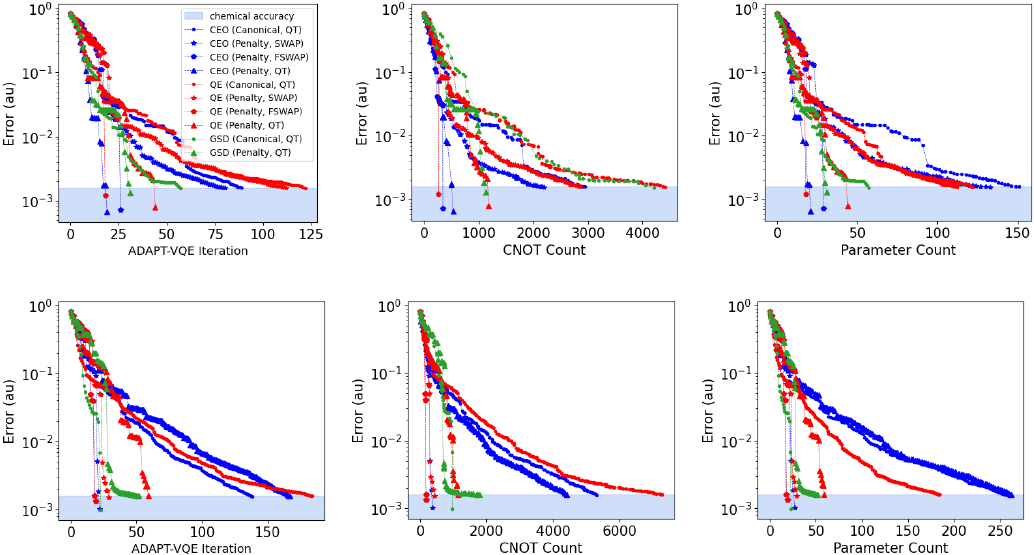}
    \caption{ADAPT-VQE error \textit{versus} iteration number, \texttt{CNOT} count for LNN connectivity, and parameter count for linear (top) and triangular (bottom) H$_6$ (12 qubits) with interatomic distance 3 \AA{}. The pools considered are CEO \cite{ramoaReducingResourcesRequired2024}, QE \cite{yordanovQubitexcitationbasedAdaptiveVariational2021} and GSD \cite{grimsleyAdaptiveVariationalAlgorithm2019}. For each of these pools, we consider Co-ADAPT-VQE with the \texttt{SWAP}, \texttt{FSWAP}, and Qiskit-transpiler (QT) -based penalties. The canonical version of ADAPT-VQE transpiled to the same (LNN) connectivity is also included for comparison. For the case of CEO and QE pools, the \texttt{FSWAP}-based version of Co-ADAPT-VQE implements the fermionic version of these operators, which is obtained by exchanging \texttt{SWAP} gates for \texttt{FSWAP}. This exchange amounts to adding Jordan-Wigner anticommutation strings to the corresponding generators. All algorithms terminate at chemical accuracy. In the bottom plots, the QE and CEO with \texttt{FSWAP}-based penalties overlap. Note that the GSD pool has no associated curve for \texttt{SWAP}- and \texttt{FSWAP}-based penalties. This is because the QE pool with \texttt{FSWAP} gates already implements GSD operators (with \texttt{FSWAP}s + QE circuit constructions), and implementing a fermionic operator using \texttt{SWAP} rather than \texttt{FSWAP} gates results in unnecessary \texttt{CNOT} gate overheads. }
    \label{fig:h6}
\end{figure*}

Figure \ref{fig:h6} compares the regular ADAPT-VQE transpiled to LNN connectivity with Co-ADAPT-VQE for three different pools. All three variants of Co-ADAPT-VQE for \texttt{CNOT} minimization defined in the previous section are considered. We consider a H$_6$ molecule as a toy model for a strongly correlated system. Linear and triangular geometries are both considered to analyze the performance of the algorithm as correlations become stronger. Since the triangular H$_6$ molecule exhibits spin-frustration, it is particularly difficult to simulate.

We observe that Co-ADAPT-VQE with the \texttt{FSWAP}-based penalty is the best performer for both molecules. This variant is able to decrease the \texttt{CNOT} count from over 4000 to roughly 300 for linear H$_6$, and from over 7000 to roughly 200 for triangular H$_6$. It is interesting to note that the impact is higher for the more strongly correlated system, which we attribute to the improved capability (as compared to qubit excitations) of fermionic excitations, which feature the Jordan-Wigner anticommutation string, to represent strongly correlated ground states. This effect has been noted previously in the literature \cite{yordanovQubitexcitationbasedAdaptiveVariational2021}, although previous work did not take advantage of the hardware efficiency of \texttt{FSWAP} networks for implementation for LNN connectivity. The instances of Co-ADAPT-VQE with \texttt{SWAP}- and Qiskit-transpiler-based penalties are also capable of reducing the CNOT counts of the circuits, although the reduction is not as significant as for the \texttt{FSWAP}-based protocol. The iteration and parameter counts required to reach chemical accuracy are actually reduced for the best performing variants of Co-ADAPT-VQE as compared to regular ADAPT-VQE; hence, we do not expect our algorithm to imply an increase in measurement costs from gradient measurement rounds and optimizations, on the contrary.

\begin{figure*}
    \centering
    \includegraphics[width=0.95\linewidth]{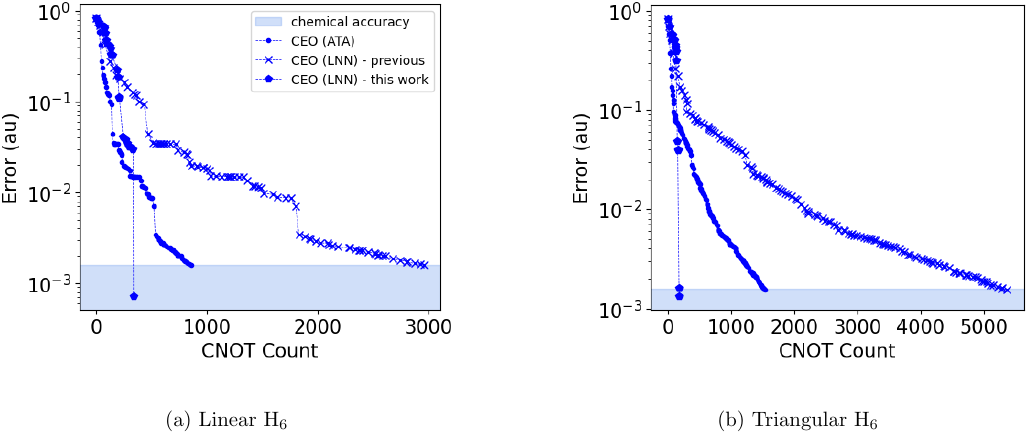}
    \caption{Error \textit{versus} \texttt{CNOT} count for CEO-ADAPT-VQE for ATA connectivity as compared to LNN connectivity, for H$_6$ at interatomic distance 3 \AA{}. The LNN \texttt{CNOT} counts are obtained in two ways: By transpiling the ATA circuit from the canonical CEO-ADAPT-VQE (the best performing pool in the literature), and by using our Co-ADAPT-VQE protocol with the CEO pool and the \texttt{FSWAP}-based penalty. All algorithms terminate at chemical accuracy.}
    \label{fig:ceo_ata_lnn}
\end{figure*}

Considering that most work on ADAPT-VQE has been done considering ATA connectivities \cite{tangQubitADAPTVQEAdaptiveAlgorithm2021,yordanovQubitexcitationbasedAdaptiveVariational2021,ramoaReducingResourcesRequired2024}, it is relevant to compare the ATA and LNN gate counts. We can obtain the LNN ans\"atze by transpiling the canonical ADAPT-VQE to LNN connectivity --- the standard method prior to this work --- or by using our co-designed version of the algorithm. Figure \ref{fig:ceo_ata_lnn} compares the \texttt{CNOT} count of both approaches with the ATA ansatz for the same molecules as Fig.~\ref{fig:h6}.

Surprisingly, our Co-ADAPT-VQE protocol is able to \textit{decrease} the CNOT count as compared to the original ATA ansatz, despite the strong restriction on connectivity (with each qubit in the connectivity graph sharing an edge with only 2 others for LNN, as compared to 14 for ATA). We attribute this to two things: first, the aforementioned fact that the Jordan-Wigner anticommutation improves the convergence of the algorithm for strongly correlated systems; and second, the fact that in the original algorithm, operators with a lower \texttt{CNOT} count are not favored in any way. While the impact of this is more pronounced in restricted connectivity, operators also have varying \texttt{CNOT} counts for ATA connectivity; for example, a single qubit excitation requires only 2 \texttt{CNOT}s, while a double qubit excitation requires 13. This brings the question of whether Co-ADAPT-VQE may be used to reduce \texttt{CNOT} counts for ATA connectivity as well. In App.~\ref{ap:ata_penalty}, we confirm that this is indeed a possibility. While the impact is not as dramatic as for the LNN case, we still observe a significant (roughly 20\%) decrease in the gate count.

\begin{figure}
    \centering
    \includegraphics[width=0.95\linewidth]{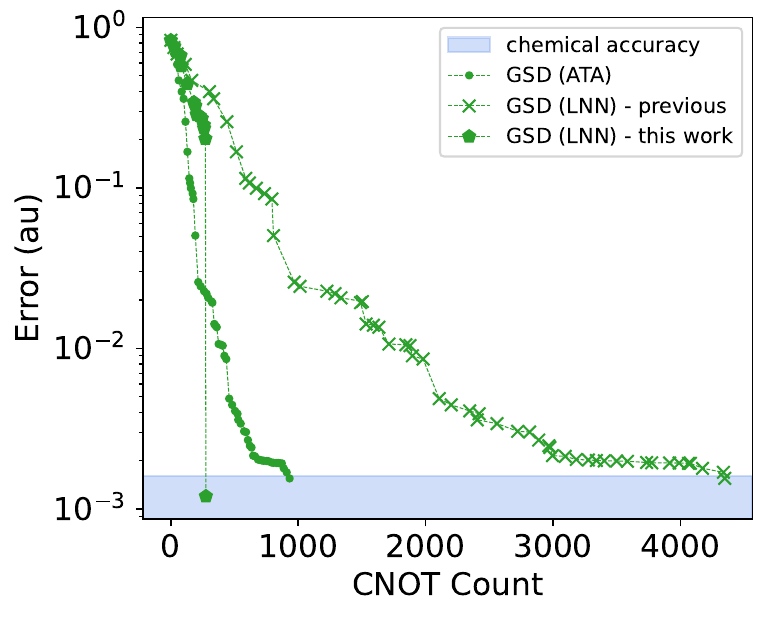}
    \caption{Error \textit{versus} \texttt{CNOT} count for GSD-ADAPT-VQE for ATA connectivity as compared to LNN connectivity. The details are the same as for Fig.~\ref{fig:ceo_ata_lnn}(a), with the only difference being the pool (GSD vs CEO). In particular, the curve associated with this work corresponds to Co-ADAPT-VQE with the GSD pool and the \texttt{FSWAP}-based penalty, such that operators are implemented as cascades of \texttt{FSWAP} gates and qubit excitation circuits.}
    \label{fig:gsd_ata_lnn}
\end{figure}

To generalize Fig.~\ref{fig:ceo_ata_lnn} to other pools, we present similar results for the GSD pool in Fig.~\ref{fig:gsd_ata_lnn}. Similarly to the case of the CEO pool, we observe that not only does Co-ADAPT-VQE result in a LNN-connectivity-restricted circuit with an order of magnitude less \texttt{CNOT}s than the canonical ADAPT-VQE ansatz after transpilation, but it also reduces this gate count as compared to the original circuit with no connecitivity restrictions.

\begin{figure*}
    \centering
    \includegraphics[width=0.8\linewidth]{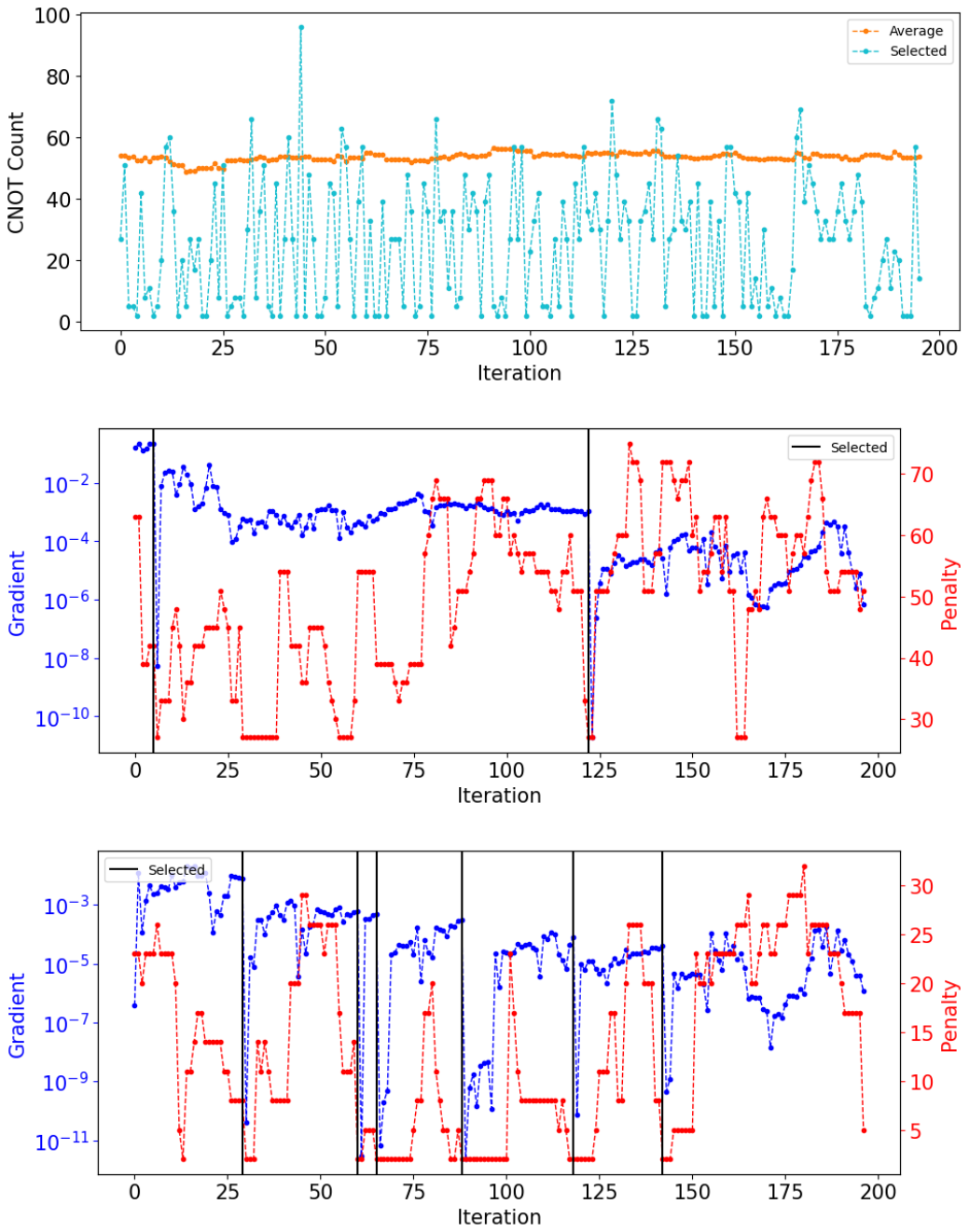}
    \caption{Evolution of the penalty and operator gradients for the Co-ADAPT-VQE algorithm. We consider the \texttt{FSWAP}-based variant with the CEO pool, applied to the linear H$_6$ molecule at interatomic distance 3 \AA{}. The top panel compares the average penalty (equivalently, \texttt{CNOT} count) with the penalty of the selected operator for each iteration. The middle panel shows the evolution of the gradient norm and the penalty for a particular double excitation, and the bottom panel shows the same results for a particular single excitation. Black vertical bars indicate iterations in which the operator was chosen.}
    \label{fig:penalty}
\end{figure*}

In Fig.~\ref{fig:penalty} we delve into data concerning the evolution of the penalty as well as operator gradients to showcase the inner workings of the algorithm. The first plot compares, for each iteration, the average \texttt{CNOT} count (across all pool operators) with the \texttt{CNOT} count of the selected operator. We observe that, as expected, Co-ADAPT-VQE tends to select operators requiring fewer \texttt{CNOT}s than average. Interestingly, this is not always the case; there are some iterations where the \texttt{CNOT} count of the selected operator is significantly higher than average. This shows that the algorithm is still able to select operators with higher \texttt{CNOT} counts, provided that their gradient magnitude is high enough to offset the penalty.

In the second and third plots, we observe the evolution of the gradient magnitude and the penalty (\texttt{CNOT} count) for a double and a single excitation, respectively. The single excitation is selected more often than the double excitation, which is typical behavior, considering single excitations can typically be implemented with fewer gates. In both cases, we observe that operators are typically selected after a big drop in the penalty arising from a favorable rearrangement of fermionic modes in the iteration before. In general, we observe that the gradient magnitude stays relatively constant until a sharp drop in the penalty occurs, at which point the operator is selected. This behavior illustrates that the algorithm favors hardware efficiency by waiting until an operator can be conveniently implemented to effectively add it to the ansatz. 

Finally, we remark on a few key points concerning our algorithm and results. First, while Co-ADAPT-VQE is ideal for chemically accurate results, we expect its impact to be less significant for extremely low errors. This is because, as we get very close to the ground energy, there are very few operators that can actually rotate the state closer to the target; hence, selecting the top operator becomes more crucial, which the application of the penalty prevents. We illustrate this in App.~\ref{ap:full_acc}, where we show that while Co-ADAPT-VQE is still able to improve upon the canonical algorithm, the impact is not as remarkable. To counterbalance this, we could modify the penalty such that it is dynamically adapted along the iterations. An interesting possibility is to add a damping term that decreases the impact of the penalty towards later iterations. This term could be a function of the iteration number, the energy changes, the gradients, and/or other available metrics. Second, the choice of Eq.~\eqref{eq:penalty} for penalizing the gradient is not unique. In fact, we can easily conceive alternative formulas for penalizing the gradient, such as the one we obtain by replacing $p$ (the penalty) with a power of it. We study this possibility in App.~\ref{ap:pen_scaling}. Third, while we only considered H$_6$ molecules in different geometries, the algorithm may be straightforwardly applied to other systems. However, for very small or weakly correlated systems requiring very shallow ans\"atze, standard transpilation techniques (which include reordering qubits at the beginning and end of the circuit) work well and might actually outperform some variants of Co-ADAPT-VQE. We illustrate this by applying the algorithm to LiH in App.~\ref{ap:lih}. However, we note that it is expected that larger and more strongly correlated molecules are more important targets when aiming for quantum advantage, and we expect Co-ADAPT-VQE to perform well for such systems. The fact that it leads to such large reductions in gate counts for H$_6$ in linear geometry, and even greater ones for the more strongly correlated triangular geometry, supports this conjecture.

\section{Conclusion}
\label{s:conclusion}

Due to its ability to avoid trainability issues and its shallow circuits, ADAPT-VQE is increasingly regarded as the leading near-term algorithm for state preparation; regardless, research efforts directed at improving the algorithm have mostly been agnostic to the target hardware. Gate counts in all-to-all connectivity are a standard metric for evaluating the algorithm's performance, but hardware constraints often enforce restrictions on which pairs of qubits can undergo entangling operations. While standard techniques can be used to transpile any ansatz to restricted connectivity, they incur a significant overhead in gate counts, which is counter to the objectives of a near-term algorithm. This begs the question: Can we design a state preparation algorithm that constructs a hardware-specific circuit that improves noise resilience without sacrificing accuracy?

In this work, we provided an affirmative answer to this question by introducing Co-ADAPT-VQE, an algorithm that co-designs an ansatz by including hardware-specific constraints and limitations into the criteria used to build the circuit. This was achieved by incorporating factors of choice---such as gate counts in restricted connectivity, operation error rates, etc.---into the selection criterion of ADAPT-VQE. This fertile framework can be adapted to suit any hardware platform and any particular quantum computer by favoring the addition of circuit blocks that can be conveniently implemented in the target device. Importantly, the framework's generality allows one to address whatever one desires to prioritize in a given experiment, whether that is achieving a low \texttt{CNOT} count or depth, using operations with low error rates, or building ans\"atze with lower parameter counts, among many other options.

To illustrate the effectiveness of Co-ADAPT-VQE, we created several instances of the algorithm to address a common problem: minimizing the \texttt{CNOT} counts of ans\"atze for preparing molecular ground states in linear nearest-neighbor connectivity, which is directly related to the noise-resilience of the circuits. We showed that Co-ADAPT-VQE achieves a dramatic decrease in this figure, with \texttt{CNOT} counts decreasing from over 7000 to fewer than 200 in the case of the most strongly correlated system (H$_6$ in triangular geometry). We expect the impact to become even greater for larger systems, where the path between pairs of qubits is on average longer in the LNN connectivity graph. Further, our simulations show that Co-ADAPT-VQE has a greater impact for more strongly correlated systems, suggesting that this algorithm will perform particularly well in simulating precisely those systems for which quantum computers are expected to provide an advantage.

\section*{Acknowledgments}

SEE and MR acknowledge support by Wellcome Leap as part of the Quantum for Bio Program. NJM and EB acknowledge support from the US Department of Energy (Grant No. DE-SC0025430). MR acknowledges support from FCT (Fundação para a Ciência e a Tecnologia, I.P.), under PhD research scholarship 2022.12333.BD. This work is funded in part by national funds through FCT, under the support UID/50014/2025 (https://doi.org/10.54499/UID/50014/2025).

\appendix

\onecolumngrid
\begin{center}
\section{Co-ADAPT-VQE for ATA connectivity}
\label{ap:ata_penalty}

\end{center}
\vspace{1em}
\begin{figure*}
    \centering
    \includegraphics[width=0.95\linewidth]{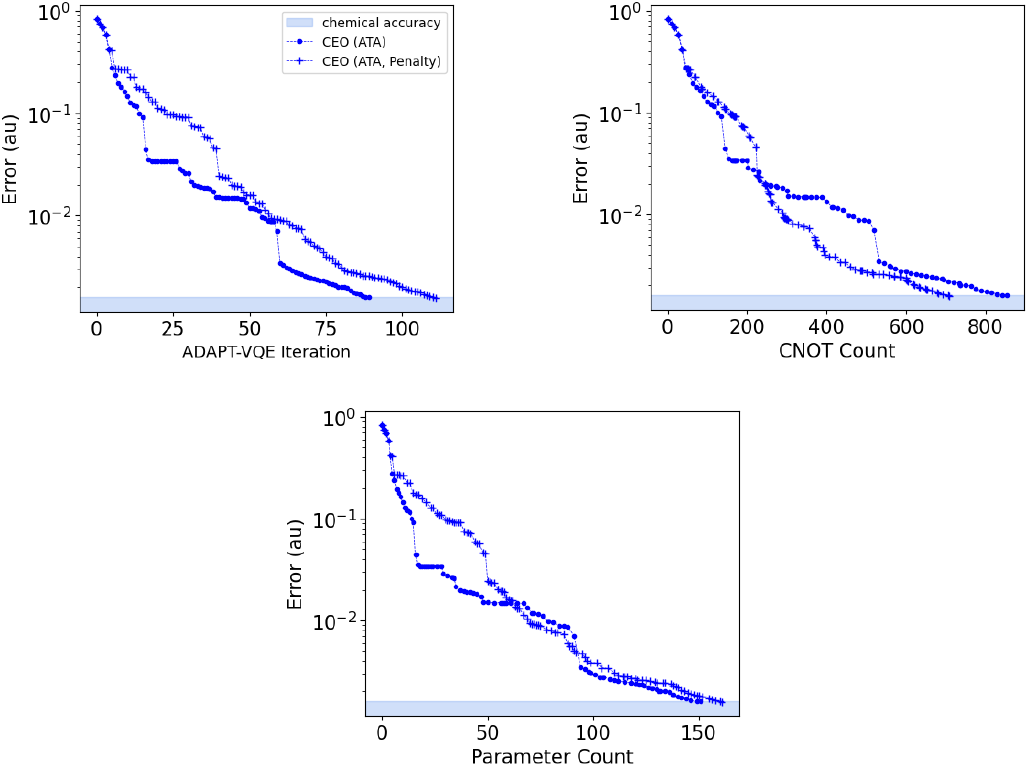}
    \caption{Error \textit{versus} ADAPT-VQE iteration, \texttt{CNOT} count, and parameter count for linear H$_6$ at interatomic distance 3 \AA{}. Two variants are considered: CEO-ADAPT-VQE and the co-designed version, where the gradients are divided by the corresponding \texttt{CNOT} count for ATA connectivity. Both algorithms terminate when chemical accuracy is reached.}
    \label{fig:ata_penalty}
\end{figure*}
\vspace{1em}

In the main text, we analyzed the impact of Co-ADAPT-VQE in producing state preparation circuits for LNN connectivity with low \texttt{CNOT} counts. We remark that while the impact will be greater in such a setting, due to the overhead incurred by the transpilation to restricted connectivity, all operator pools proposed in the literature include operators with varying \texttt{CNOT} counts. Hence, Co-ADAPT-VQE can also be applied to reduce gate counts for ATA connectivity. 

We illustrate this in Fig.~\ref{fig:ata_penalty} for the linear H$_6$ molecule. We observe that while Co-ADAPT-VQE requires slightly more iterations to converge to chemical accuracy and terminates with slightly more parameters, it does succeed in decreasing the \texttt{CNOT} count by roughly 20\%. This tradeoff between gate counts and parameters is usually favorable, since the size of the circuit determines whether or not it can be viably executed to reasonable accuracy. 

We note that the purpose of this appendix is to analyze the impact of the penalty-based selection criterion on ADAPT-VQE for ATA connectivity. Parallel to this, we observe that the leading versions of Co-ADAPT-VQE improve the energy convergence by using \texttt{FSWAP}s in the operator routing. While this choice may at first be motivated by the lower number of \texttt{CNOTs} necessary to implement an \texttt{FSWAP} gate as compared to a \texttt{SWAP} gate, this additionally contributes to convergence due to the favorable action of the Jordan-Wigner $Z$ strings for strongly correlated systems \cite{yordanovQubitexcitationbasedAdaptiveVariational2021}.

\clearpage
\onecolumngrid
\begin{center}
\section{Co-ADAPT-VQE for Higher Accuracy}
\label{ap:full_acc}
\end{center}
\vspace{1em}
\begin{figure*}
    \centering
    \includegraphics[width=0.95\linewidth]{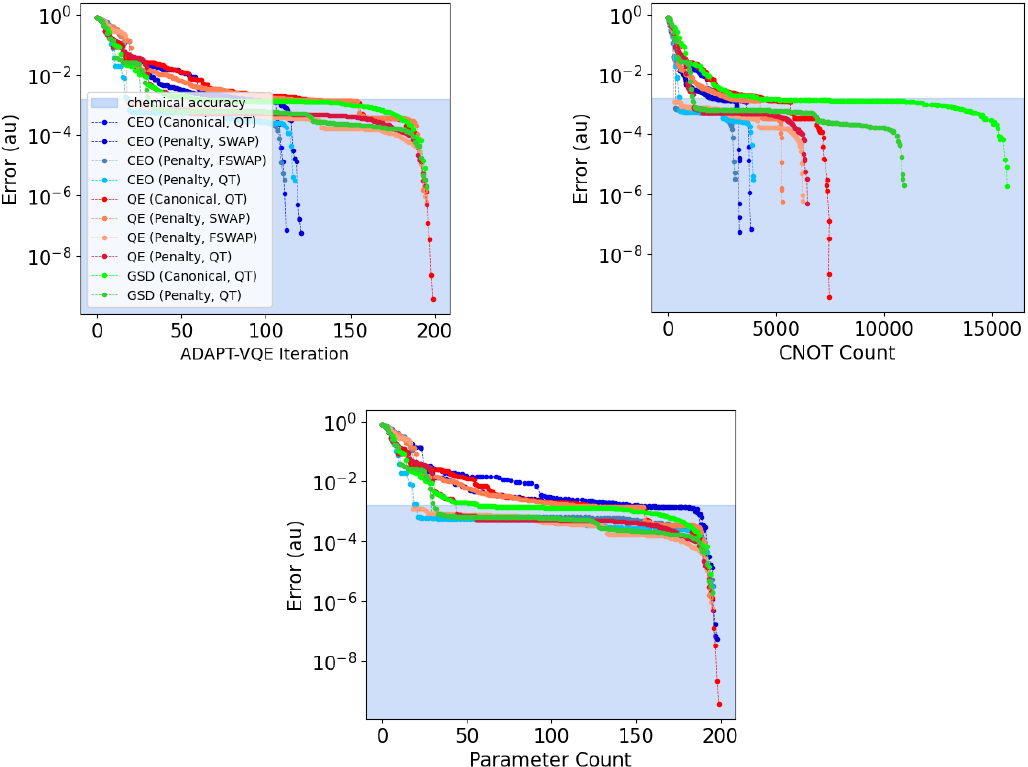}
    \caption{ADAPT-VQE error \textit{versus} iteration number, \texttt{CNOT} count for LNN connectivity, and parameter count for linear H$_6$ (12 qubits) with interatomic distance 3 \AA{}. All algorithms terminate when the gradient norm falls below $10^{-4}$.}
    \label{fig:full_acc}
\end{figure*}
\vspace{1em}

Throughout this work, we focused on the gate and parameter counts required to reach chemically-accurate energies. This is the most important regime, since it offers experimentally useful results without aggravating the high measurement costs of molecular ground state preparation \cite{gonthierMeasurementsRoadblockNearterm2022}. However, for completeness, we study a higher accuracy regime in Fig.~\ref{fig:full_acc}, with energy errors going as low as $10^{-10}$ Hartree. We consider the same variants of ADAPT-VQE as in Fig.~\ref{fig:h6} of the main text.

We observe that while Co-ADAPT-VQE is still able to offer a decreased gate count with respect to the canonical version of the algorithm, the improvement is not as significant as in the chemically-accurate regime with errors around $10^{-3}$ Hartree. We attribute this to the fact that once we reach very low errors, we are very limited in the range of operators we can choose to approach the ground state, as very specific excitations will be required to fine-tune the final wavefunction.

\clearpage
\onecolumngrid
\begin{center}
\section{Penalty Scaling}
\label{ap:pen_scaling}
\end{center}
\vspace{1em}
\begin{figure*}
    \centering
    \includegraphics[width=0.95\linewidth]{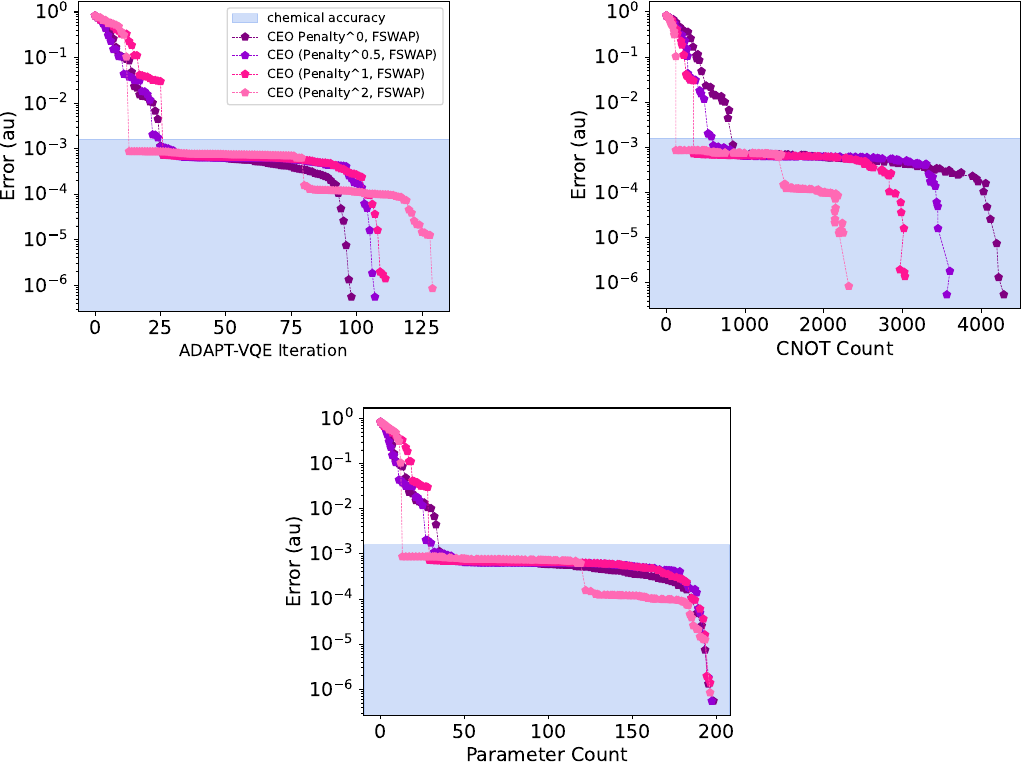}
    \caption{Impact of scaling the penalty via varying exponentials on the performance of Co-CEO-ADAPT-VQE for the linear H$_6$ molecule. A darker curve corresponds to a lower exponent.}
    \label{fig:penalty_scaling}
\end{figure*}
\vspace{1em}

In the proposal of the Co-ADAPT-VQE algorithm introduced in the main text, we modify the selection criterion by dividing the gradient by a penalty of choice (Eq.~\eqref{eq:penalty}). However, this is not the only way of incorporating hardware characteristics into the selection criterion. 

To exemplify alternative protocols, we make the penalty weaker or stronger by modifying the equation defining the selection criterion to
\begin{equation}
    \frac{\pdv{E^{(n)}}{\theta_i}\Bigg|_{\theta_i=0}}{p^k} = \frac{\bra{\psi^{(n-1)}}\left[\hat{H},\hat{A}_i\right]\ket{\psi^{(n-1)}}}{p^k},
    \label{eq:penalty_scaled}
\end{equation}
for various values of the hyperparameter $k$. In particular, we take $k=0$ (no penalty), $k=0.5$ (penalty weaker than in the main text), $k=1$ (penalty as in the main text), and $k=2$ (penalty stronger than in the main text). Figure \ref{fig:penalty_scaling} shows the results of the penalty scaling. As expected, we observe that stronger penalties lead to lower \texttt{CNOT} counts while requiring more iterations. 

\clearpage
\onecolumngrid
\begin{center}
\section{Results for LiH}
\label{ap:lih}
\end{center}
\vspace{1em}
\begin{figure*}
    \centering
    \includegraphics[width=0.95\linewidth]{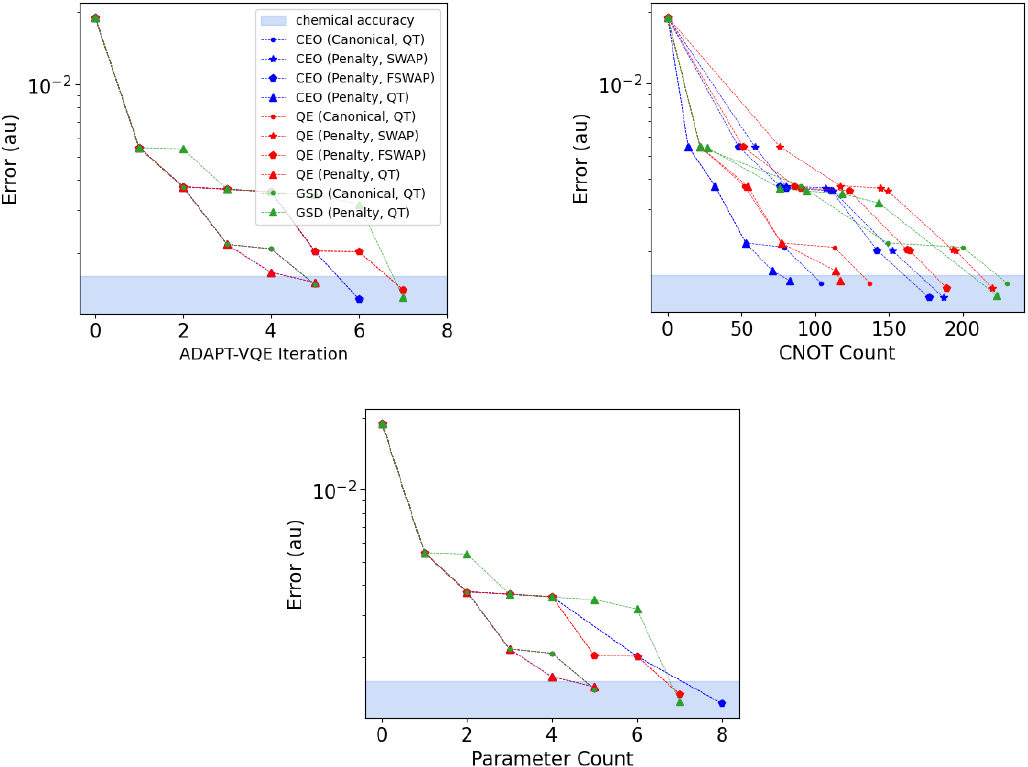}
    \caption{Error \textit{versus} the iteration number, LNN \texttt{CNOT} count, and parameter count for multiple variants of ADAPT-VQE applied to the LiH molecule (12 qubits) at interatomic distance 1.5 \AA{}. We consider three different pools (CEO, QE, and GSD), three different variants of Co-ADAPT-VQE (\texttt{SWAP}-based, \texttt{FSWAP}-based, and Qiskit-transpiler-based), and the canonical ADAPT-VQE algorithm.}
    \label{fig:lih}
\end{figure*}
\vspace{1em}

In the main text, we focused on two strongly correlated molecules: linear and triangular H$_6$. For the sake of completeness, we include results for the LiH molecule, an easier to simulate (less correlated) 12-qubit system.

The resulting \texttt{CNOT} and parameter counts are plotted in Fig.~\ref{fig:lih}. We observe that while some variants of Co-ADAPT-VQE decrease the \texttt{CNOT} counts with respect to the canonical ADAPT-VQE algorithm transpiled to LNN connectivity, that is not always the case, and the improvement is less significant than for the systems studied in the main text.

We attribute this difference to the fact that for very shallow ans\"atze, the transpiler achieves a significant improvement of the circuit by reordering the initial and final qubit assignments. For example, regardless of the first excitation that is chosen, we can always assume that it acts on adjacent qubits at the start of the algorithm, since we can redefine the initial qubit order for that to be the case. Recall that we apply Qiskit's transpilation with maximum optimization to the original ADAPT-VQE circuit; hence, these transpilation techniques that work well for shallow circuits will yield particularly good results for molecules such as LiH, where the ansatz needs no more than a few operators.

However, we note that such shallow ans\"atze are within the realm of classical simulability, and that larger / more strongly correlated systems are bound to be the relevant use cases of quantum computers. For those cases, as we show in the main text, we have deeper ans\"atze which can benefit the most from Co-ADAPT-VQE.

\FloatBarrier

\bibliography{main.bib}

\end{document}